\documentclass[journal]{IEEEtran}

\usepackage{amsmath}
\usepackage{amsfonts,amssymb}
\usepackage{graphicx}
\allowdisplaybreaks[4]
\usepackage{xcolor}
\definecolor{deepred}{RGB}{205,38,38}
\usepackage{algorithm}
\usepackage{algorithmic}
\usepackage{amsthm}

\usepackage{lineno,hyperref}
\modulolinenumbers[5]
\usepackage{bm}
\DeclareMathOperator*{\argmin}{arg\,min}
\usepackage{multirow}
\usepackage{array}
\usepackage{booktabs}
\usepackage{lipsum}
\usepackage{xcolor}
\usepackage{etoolbox}
\ifCLASSINFOpdf
\else
\fi
\ifCLASSOPTIONcompsoc
  \usepackage[caption=false,font=normalsize,labelfont=sf,textfont=sf]{subfig}
\else
  \usepackage[caption=false,font=footnotesize]{subfig}
\fi

\graphicspath{{./FiguresJournal/}} 

\begin{document}

\title{Automatic Self-Adaptive Local Voltage Control Under Limited Reactive Power} 
\author{Rui Cheng,~\IEEEmembership{Graduate Student Member,~IEEE}, Naihao Shi,~\IEEEmembership{Graduate Student Member,~IEEE}, Salish~Maharjan,~\IEEEmembership{Member,~IEEE}, Zhaoyu Wang,~\IEEEmembership{Senior Member,~IEEE}%
\thanks{This work was supported in part by the U.S. Department of Energy Wind Energy Technologies Office under Grant DE-EE0008956, and in part by the National Science Foundation under ECCS 1929975 (Corresponding author: Zhaoyu Wang).} 
\thanks{Rui Cheng, Naihao Shi, Salish Maharjan and Zhaoyu Wang are with the Department of Electrical and Computer Engineering, Iowa State University, Ames, IA 50011 USA (e-mail: ruicheng@iastate.edu; snh0812@iastate.edu; salish@iastate.edu; wzy@iastate.edu).}
}

\maketitle
\begin{abstract} 
The increasing proliferation of distributed energy resources has posed new challenges to Volt/VAr control problems in distribution networks. To this end, this paper proposes an automatic self-adaptive local voltage control (ASALVC) by locally controlling VAr outputs of distributed energy resources. In this ASALVC strategy, each bus agent can locally and dynamically adjust its voltage droop function in accordance with time-varying system changes. The voltage droop function is associated with the bus-specific time-varying slope and intercept, which can be locally updated, merely based on local voltage measurements, without requiring communication. Stability, convergence, and optimality properties of this local voltage control are analytically established. {In addition, the online implementation of ASALVC is further proposed to address the real-time system changes by adjusting VAr outputs of DERs online.}
Numerical test cases are performed to validate and demonstrate the effectiveness and superiority of ASALVC.
\end{abstract}


\section{Introduction}
\label{sec:Intro}
\IEEEPARstart{R}ecent years have seen the increasing deployment and penetration of renewable energy resources, such as photovoltaic (PV) generators and wind, in power systems, which has led to  over-/under-voltage problems due to the intermittent and volatile nature of renewable energy resources. Volt/VAr Control (VVC) strategies have shown a great capability to effectively resolve those voltage problems by controlling VAr outputs owing to the rapid development of inverter-based technologies for distributed energy resources (DERs) \cite{KT}. 

In the past decades, VVC strategies have been extensively and widely studied by researchers and practitioners.  In general, it can be roughly divided into three main categories: centralized voltage control, distributed voltage control, {and local voltage control}.

Centralized voltage control (\cite{MF}-\cite{CVR}) collects all the required information, such as network and load parameters, and then performs a central computation to solve the corresponding optimization and control problems. However, it usually suffers from large amounts of computation time, considerable communication overload, and privacy problems, hindering scalability. 

Rather than collecting all problem parameters and performing a central calculation, distributed voltage control is computed by many agents that obtain certain parameters via coordinating communication \cite{OV2}. According to the coordinating communication infrastructure, it can be further divided into hierarchical voltage control, where a central agent communicates with other agents in a hierarchical manner, and decentralized voltage control, where each agent communicates with its neighbors, but there is not a central agent. For example,  hierarchical voltage control schemes, based on the Alternating Direction Method of Multipliers (ADMM) or projected Newton method, are applied to coordinate electric vehicle charging schedules, wind turbines, {photovoltaic} inverters in \cite{XZ}-\cite{RC}, respectively. Different network-constrained ADMM-based decentralized voltage control strategies are proposed in \cite{BAR}-\cite{PS}, relying on the communication between neighboring buses. Furthermore, some advanced online decentralized voltage control algorithms are developed in \cite{HJL}-\cite{GQ}, where the real-time measurements are utilized to determine the control solution. 

Compared to centralized voltage control and distributed voltage control, local voltage control typically only relies on local information without requiring communication, rendering itself a more practical and scalable implementation. The traditional droop control \cite{AdvancedVVC,MF2}, as advocated by IEEE 1547-2018 Standard \cite{1547-2018}, is one of the most common and popular local voltage control, which actively adjusts the VAr output as a function of voltage following a given `Volt-VAr' piecewise linear characteristic. However, as shown in \cite{MF2}, the droop slope in the traditional droop control needs to be small enough to guarantee system stability. Moreover, the work \cite{NL} shows such {a} traditional droop control is not able to maintain a feasible voltage profile under certain circumstances. A modified `delayed' droop control is proposed in \cite{PJ} to improve the stability performance, but it is unclear how to determine the delay parameter to optimally balance the stability performance and convergence speed. The works \cite{PJ}-\cite{GC} provide stability analysis, but all of them lack the optimality analysis and system-wide performance  characterization resulting from the implementation of local voltage control. The works \cite{YG}-\cite{HZ} formulate the local voltage control as optimization problems, exhibiting a better stability performance than the traditional and delayed droop control, where rigorous stability, convergence, and optimality analyses are provided.
The studies \cite{YG}-\cite{HZ} are both based on the (scaled/classical) gradient projection (GP) method, which can be regarded as one type of modified voltage droop control with a constant slope and a time-varying intercept. However, the convergence rate of GP is relatively slow \cite{BertsekasNP,ProjectedGradient}, which is typically characterized by $O(1/k)$ ($k$ is the iteration number), indicating  relatively weak tracking capabilities to follow system variations. Moreover, the constant slope in \cite{YG}-\cite{HZ} limits the diversity and flexibility of local voltage control to some degree.

To this end, this paper proposes an automatic self-adaptive local voltage control (ASALVC) to solve the VVC problem with the goal of mitigating the voltage deviations across distribution networks by locally controlling VAr outputs of DERs. This VVC problem is formulated as an optimization problem,  which is first solved by a generalized fast gradient method (GFGM) \cite{AB,ProximalGradient}. Interestingly, the GFGM iterations naturally decouple into  \textit{communication-free} local updates, which can be reinterpreted as our proposed local voltage control
strategy, by properly choosing and designing parameters. Compared with existing studies,  the main contributions of this study are as follows:
\begin{itemize}
    \item 
    This local voltage control is \textit{automatic self-adaptive}, allowing each bus agent to locally and dynamically adjust its voltage droop function in accordance with time-varying system changes. This voltage droop function is associated with both \textit{the bus-specific time-varying slope and intercept}, significantly increasing the diversity and flexibility of local voltage control. 
    \item The time-varying slope and intercept are locally and intelligently updated by each bus agent merely based on its local voltage measurements without requiring communication, where  \textit{the closed-form expressions} of the bus-specific time-varying slope and intercept are analytically explored and presented. 
    \item This automatic self-adaptive local voltage control exhibits \textit{an accelerated convergence rate} both theoretically and practically, characterized by $O(1/k^2)$, in static scenarios, indicating \textit{a better tracking capability} to follow time-varying changes in dynamic scenarios. \textit{Stability, convergence, and optimality properties} of this self-adaptive local voltage control are first analytically established and then demonstrated by means of numerical test cases.
\end{itemize}
{Remaining sections are organized as follows. The network modeling and problem statement are discussed in Section~\ref{sec:ProblemStatement}. The GFGM-based VVC strategy is described in Section~\ref{sec:GFGM}. Section~\ref{sec:ALVC} carefully explains the transition from the GFGM-based VVC strategy to this ASALVC by properly choosing and designing parameters. Both the offline and online implementation of this ASALVC are demonstrated in Section~\ref{sec:ALVC}. Section~\ref{sec:Case} reports numerical test cases, and Section~\ref{sec:Conclusion} discusses ongoing and planned future studies.}
\begin{table}[t]
    \centering
    \caption{Nomenclature: Operator}
    \label{tab:Nomenclature}
    \begin{tabular}{ll}
    \hline
        $<\bm{x},\bm{y}>$ & It denotes $\bm{x}^T\bm{y}$\\
        $<\bm{x},\bm{y}>_{\bm{L}}$& It denotes $\bm{x}^T\bm{L}\bm{y}$\\
        $||\bm{x}||_{\bm{L}}^2$ & It denotes $\bm{x}^T\bm{L}\bm{x}$ \\
        $\bm{X}\succeq\bm{Y}$ & $\bm{X}-\bm{Y}$ is semi-positive definite.\\
        $\bm{X}=\text{diag}(x_{1},...,x_{n})$ &  A square diagonal matrix with the elements \\&$x_{1},...,x_{n}$ on the main diagonal.\\
        $\sigma_{\rm min}(\cdot)$& It denotes the smallest eigenvalue.\\
    \hline
    \end{tabular}
\end{table}
\section{{Network Modeling and Problem Statement}}
\label{sec:ProblemStatement}
Consider a radial distribution network with $N+1$ buses. Let $\{0\}\bigcup\mathcal{N}$ denote the bus set, where ${\mathcal{N}} = \{1,2,...,N\}$. For each bus $j\in\mathcal{N}$, 
let $V_i$ denote its voltage magnitude,  $p_i$ and $q_i$ denote its real and reactive power injections. Let $b^p(j)\in\{0\}\cup{\mathcal{N}}$ denote the bus immediately preceding bus $j$ along the radial distribution network, $\mathcal{L}=\{\ell_j=(i,j)| i=b^p(j),j\in\mathcal{N}\}$ denote the line segment set. For each line segment $(i,j)\in\mathcal{L}$, let $r_{ij}$ and $x_{ij}$ denote its resistance and reactance, $P_{ij}$ and $Q_{ij}$ denote the real and reactive power flows from bus $i$ to $j$, respectively.  Also, let $\mathcal{N}_j$ denote the set of all buses located strictly after bus $j$ along the radial network. The branch flow model \cite{DistFlow} to model this radial distribution network flow is given for $\forall{(i,j)}\in\mathcal{L}$ as follows:
\begin{subequations}\label{eq:nonlinear}
\begin{align}
P_{ij}-\sum_{k\in\mathcal{N}_j}P_{jk}&=-p_j+r_{ij}\frac{P_{ij}^2+Q_{ij}^2}{V_i^2}\\
Q_{ij}-\sum_{k\in\mathcal{N}_j}Q_{jk}&=-q_j+x_{ij}\frac{P_{ij}^2+Q_{ij}^2}{V_i^2}\\
\nonumber{V}_i^2-V_j^2&=2(r_{ij}P_{ij}+x_{ij}Q_{ij})\\&-(r_{ij}^2+x_{ij}^2)\frac{P_{ij}^2+Q_{ij}^2}{V_i^2}
\end{align}
\end{subequations}
And we further define the column vectors $\bm{V}=[V_i]_{i\in\mathcal{N}}$, $\bm{p}=[p_i]_{i\in\mathcal{N}}$, $\bm{q}=[q_i]_{i\in\mathcal{N}}$, $\bm{P}=[P_{b^p(j)j}]_{(b^p(j),j)\in\mathcal{L}}$, $\bm{Q}=[Q_{b^p(j)j}]_{(b^p(j),j)\in\mathcal{L}}$.\footnote{The real and reactive power flows over line segments $\ell_j$ are sorted in accordance with the ordering of these line segments from small to large $j$.The bus voltage magnitudes and real/reactive power injections are sorted in accordance with the ordering of these buses from small to large $j$.} 
Before rigorously formulating the VVC problem, we separate $\bm{q}=\bm{q}^g-\bm{q}^{c}$ into two parts, i.e., $\bm{q}^g$ and $\bm{q}^c$, where $\bm{q}^g$, $\bm{q}^c$ denote the reactive power contributed by DERs and any other load reactive power consumption, respectively. The nonlinear power flow relationships existing in (\ref{eq:nonlinear}) are compactly expressed as follows:
\begin{equation}
    \bm{V}=h(\bm{q}^g,\bm{d})
\end{equation}
where $\bm{d}=\{\bm{q}^c,\bm{p}\}$. The VVC problem, based on the nonlinear power flow, aims to mitigate the voltage deviations by controlling VAr outputs of DERs. It is represented as follows: 
\begin{subequations}\label{eq:NonlinearlVoltageProb}
\begin{align}
    \min_{\bm{q}^g} &~m(\bm{q}^g)=\frac{1}{2}||h(\bm{q}^g,\bm{d})-\bm{V}_r||_{\bm{\Phi}}^2\\
    \text{s.t.~}& {\bm{\underline{q}}}^g\leq\bm{q}^g\leq{\bm{\overline{q}}}^g
\end{align}
\end{subequations}
where $\bm{V}_r\in\mathbb{R}^m$ is the reference of voltage magnitude, $\bm{\Phi}$ is a symmetric positive-definite matrix, (\ref{eq:NonlinearlVoltageProb}b) denotes VAr limits for DERs. However, this VVC problem is non-convex due to the nonlinear power flow $h(\bm{q}^g,\bm{d})$, which is challenging to solve. 

To facilitate the algorithm design and theoretical analysis, the linearized distribution power flow is adopted
\footnote{It is based on two assumptions:
(1) The loss is negligible compared to the line flow;
(2) With respect to the relatively flat voltage profile, i.e., $V_i\approx{1}$, for $\forall{i}\in\mathcal{N}$, we have $V_i^2-V_j^2=2(V_i-V_j)$. {The approximation error introduced by the two assumptions is relatively small \cite{HZ}}}:
\begin{subequations}\label{eq:linear}
\begin{align}
P_{ij}-\sum_{k\in\mathcal{N}_j}P_{jk}&=-p_j,Q_{ij}-\sum_{k\in\mathcal{N}_j}Q_{jk}=-q_j\\
V_i-V_j&=r_{ij}P_{ij}+x_{ij}Q_{ij} 
\end{align}
\end{subequations}

Consider the standard matrix representation $\bm{\bar{M}}=[\bm{m}_0, \bm{M}^T]^T\in\mathbb{R}^{(N+1)\times N}$ for the incidence matrix of a radial distribution network.\footnote{{A simple numerical example illustrating the construction of $\bm{\bar{M}}$ for a radial distribution network is given in \cite[Appendix C]{WP}.} }
Based on $\bm{\bar{M}}$, this linearized distribution power flow can be compactly denoted by:
\begin{equation}\label{eq:LinDistFlow}
    \bm{V}=\bm{M}^{-T}\bm{R}\bm{M}^{-1}\bm{p}+\bm{M}^{-T}\bm{X}\bm{M}^{-1}\bm{q}-{V}_{0}\bm{M}^{-T}\bm{m}_{0}
\end{equation}
where $\bm{R}$ and $\bm{X}$ are $N\times{N}$ diagonal matrices with $j$-th diagonal entries being the resistance and reactance of $\ell_{j}$, respectively. Let $\bm{A}=\bm{M}^{-T}\bm{X}\bm{M}^{-1}$ and $\bm{V}^{par}(\bm{d})=\bm{M}^{-T}\bm{R}\bm{M}^{-1}\bm{p}-\bm{A}\bm{q}^c-{V}_{0}\bm{M}^{-T}\bm{m}_{0}$, (\ref{eq:LinDistFlow}) can be denoted by:
\begin{equation}\label{eq:DistFlow2}
     \bm{V}=h_l(\bm{q}^g,\bm{d})=\bm{A}\bm{q}^g+\bm{V}^{par}(\bm{d})
\end{equation}
We further define $f(\bm{q}^g)$ as follows:
\begin{equation}\label{eq:fq}
\begin{split}
    f(\bm{q}^g)&=\frac{1}{2}||h_l(\bm{q}^g,\bm{d})-\bm{V}_r||_{\Phi}^2\\
    &=\frac{1}{2}||\bm{A}\bm{q}^g+\bm{V}^{par}(\bm{d})-\bm{V}_r||_{\Phi}^2
\end{split}
\end{equation}
And let $g(\bm{q}^g)$ denote the indicator function of the box constraints ${\bm{\underline{q}}}^g\leq\bm{q}^g\leq{\bm{\overline{q}}}^g$.  This VVC problem\footnote{
{For this VVC problem, we do not consider the hard voltage constraint, but instead treat the voltage constraint as a soft penalty in the objective to facilitate the algorithm design, like \cite{ZT,HZ}}}, based on the linearized distribution power flow (\ref{eq:DistFlow2}), is represented as follows: 
\begin{equation}\label{eq:ProblemReform}
    \min_{\bm{q}^g} F(\bm{q}^g)=f(\bm{q}^g)+g(\bm{q}^g)
\end{equation}

Note that $\bm{M}$ is a symmetric positive-definite matrix \cite{HZ}, it follows that $\bm{A}=\bm{M}^{-T}\bm{X}\bm{M}^{-1}$ is a symmetric positive-definite matrix, indicating $f(\bm{q}^g)$ is convex. The VVC problem (\ref{eq:ProblemReform}) turns out to be a box-constrained convex program.


\medskip
\noindent
{\textbf{Remark 1:}  The linearized distribution power flow (\ref{eq:LinDistFlow}) is adopted to facilitate the algorithm design and theoretical analysis. Note that our proposed voltage control can also be applied to the nonlinear distribution power flow model (\ref{eq:nonlinear}). In our numerical case studies, we test the performance of our proposed voltage control on the nonlinear power flow model.}

\section{Voltage Control Using Generalized Fast Gradient Method}
\label{sec:GFGM}
In this section, we propose a GFGM-based VVC strategy to solve the VVC problem. The GFGM-based VVC strategy is the basis for designing automatic local self-adaptive voltage control. \textit{The GFGM-based VVC strategy can be equivalently converted into automatic local self-adaptive voltage control by designing proper parameters, which will be discussed in detail in Section \ref{sec:ALVC}.} {The stability, convergence and optimality properties of the GFGM-based VVC strategy also apply to the proposed automatic local self-adaptive voltage control.}
\subsection{GFGM-Based Volt/Var Control}
For a box-constrained convex program, it is suitable to solve it by means of GP methods, but the convergence rate of GP is relatively slow \cite{BertsekasNP,ProjectedGradient}, which is typically characterized by $O(1/k)$ ($k$ is the iteration number). 
To improve the convergence performance, we apply the generalized fast gradient method \cite{AB,ProximalGradient} to solve this VVC problem (\ref{eq:ProblemReform}) with a global rate of convergence, which is proven to be significantly better compared to traditional GP methods. 

Before applying GFGM to solve this VVC problem (\ref{eq:ProblemReform}), we first introduce the concept of the approximation model, which is defined as follows:

\noindent
\textbf{Definition: Approximation model of $F(\bm{q}^g)$.} Given a symmetric positive-definite matrix $\bm{L}$, we say $Q_{\bm{L}}(\bm{q}^g,\bm{y})$  is the quadratic approximation model of  $F(\bm{q}^g)$ at a given point $\bm{y}$ if $Q_{\bm{L}}(\bm{q}^g,\bm{y})$ satisfies:
\begin{equation}\label{eq:ApproximationModel}
\begin{split}
    &F(\bm{q}^g){\leq}Q_{\bm{L}}(\bm{q}^g,\bm{y})\\&=f(\bm{y})+<\nabla{f}(\bm{y}),\bm{q}^g-\bm{y}>+\frac{1}{2}||\bm{q}^g-\bm{y}||_{\bm{L}}^2+g(\bm{q}^g)
\end{split}
\end{equation}
where
\begin{align*}
    <\nabla{f}(\bm{y}),\bm{q}^g-\bm{y}>&=[\nabla{f}(\bm{y})]^T(\bm{q}^g-\bm{y}),{\text{ and}}\\
    ||\bm{q}^g-\bm{y}||_{\bm{L}}^2&=(\bm{q}^g-\bm{y})^T\bm{L}(\bm{q}^g-\bm{y})
\end{align*}
And let $p_{\bm{L}}(\bm{y})$ be:
\begin{equation}\label{eq:pL}
    p_{\bm{L}}(\bm{y})=\argmin_{\bm{q}^g}Q_{\bm{L}}(\bm{q}^g,y)
\end{equation}
Based on the definitions of $Q_{\bm{L}}(\bm{q}^g,\bm{y})$ and $p_{\bm{L}}(\bm{y})$, the specific steps of applying GFGM to solve this VVC problem (\ref{eq:ProblemReform}) are given in Algorithm 1: GFGM-Based VVC.

\begin{algorithm}[t]
\renewcommand{\thealgorithm}{1}\selectfont
\small
\caption{GFGM-Based VVC}
\begin{algorithmic}
\STATE \hspace{-3mm}{\bf Initialization:} Set the iteration time $k=0$, and $\gamma(1)=1$, $\bm{q}^g(0)=\bm{y}(1)=\bm{0}$.
\STATE \hspace{-3mm}{\bf For} {$k\geq1$}: Alternately update variables by the following steps (S1)-(S3) until convergence: 
\STATE\hspace{-1mm} {\bf S1:} Update $\bm{q}^g(k)$:
\begin{align*}
\bm{q}^{g}(k)=p_{\bm{L}}(\bm{y}(k))=\argmin_{\bm{q}^g}Q_{\bm{L}}(\bm{q}^g,y)
\end{align*}
\STATE\hspace{-1mm} {\bf S2:}Update $\gamma(k+1)$:
\begin{align*}
    \gamma(k+1)=\frac{1+\sqrt{1+4\gamma(k)^2}}{2}
\end{align*}
\STATE\hspace{-1mm} {\bf S3:}  Update $\bm{y}(k+1)$:
\begin{align*}
    \bm{y}(k+1)=\bm{q}^g(k)+\big[\frac{\gamma(k)-1}{\gamma(k+1)}\big]\big[\bm{q}^g(k)-\bm{q}^g(k-1)\big]
\end{align*}
\end{algorithmic}
\end{algorithm}

\subsection{Stability, Convergence and Optimality Analyses}
The stability, convergence and optimality properties of Algorithm 1: GFGM-Based VVC, are established on Propositions 1-4.

\noindent
\textbf{Proposition 1:}  Assume that $f(\bm{q}^g):\mathbb{R}^{N}\xrightarrow{}\mathbb{R}$ is convex and continuously differentiable and $\bm{L}$ is a symmetric positive-definite matrix. The condition that:
\begin{equation}\label{eq:Upperbound}
f(\bm{q}^g)\leq{f}(\bm{y})+<\nabla{f}(\bm{y}),\bm{q}^g-\bm{y}>+\frac{1}{2}||\bm{q}^g-\bm{y}||_{\bm{L}}^2
\end{equation}
holds for all $\bm{q}^g,\bm{y}\in\mathbb{R}^N$ is equivalent to that:
\begin{equation}\label{eq:Lips}
    <\nabla{f}(\bm{q}^g)-\nabla{f}(\bm{y}),\bm{q}^g-\bm{y}>\leq||\bm{q}^g-\bm{y}||_{\bm{L}}^2
\end{equation}
holds for all $\bm{q}^g,\bm{y}\in\mathbb{R}^N$.
\\
\noindent
\textbf{Proof of Proposition 1:} See Appendix A.

\smallskip

\noindent
\textbf{Proposition 2:} Suppose $F(\bm{q}^g)=f(\bm{q}^g)+g(\bm{q}^g)$ satisfies the following conditions:
\begin{itemize}
    \item \textbf{[P2.A]} $g(\bm{q}^g)$ is a convex function which may not be differentiable.  
    \item \textbf{[P2.B]} $f(\bm{q}^g)$ is convex and continuously differentiable.
    \item \textbf{[P2.C]} $Q_{\bm{L}}(\bm{q}^g,y)$ is the quadratic approximation model of $F(\bm{q}^g)$
\end{itemize}
Then the sequence $\{\bm{q}^g(k)\}$, generated by Algorithm 1: GFGM-Based VVC, satisfies:
\begin{equation}\label{eq:superlinear}
    F(\bm{q}^g(k))-F(\bm{q}^{g\ast})\leq\frac{2||\bm{q}^g(0)-\bm{q}^{g\ast}||_{\bm{L}}^2}{(k+1)^2}, \forall{k}\geq{1}
\end{equation}
where $\bm{q}^{g\ast}$ is the optimal solution of (\ref{eq:ProblemReform}),  
\\
\noindent
                \textbf{Proof of Proposition 2:} Proposition 2 can be easily proved by replacing ${L}<\cdot,\cdot>$ and $L||\cdot||^2$ in the proofs in \cite{AB} with $<\cdot,\cdot>_{\bm{L}}$ and $||\cdot||_{\bm{L}}^2$. Q.E.D.

{With respect to Proposition 2, as conditions [P2.A]-[P2.C] hold, it shows Algorithm 1 GFGM-Based VVC can achieve a convergence rate no worse than $O(1/(k+1)^2)$, it exhibits a fast  convergence rate compared to GP methods with the convergence rate $O(1/k)$.}
Note that [P2.A] holds since the indicator function of the convex set ${\bm{\underline{q}}}^g\leq\bm{q}^g\leq{\bm{\overline{q}}}^g$ is a convex function. Additionally, $\nabla^2{f}(\bm{q}^g)=\bm{A}\Phi\bm{A}$ is positive definite as $\Phi$ and $\bm{A}$ are both symmetric positive definite matrices, indicating [P2.B] holds. \textit{\textbf{One remaining challenge is [P2.C].}} Note that from Proposition 1, we know that as long as $\bm{L}$ satisfies (\ref{eq:Lips}), then [P2.C] will hold. Thus, to satisfy [P2.C], the symmetric positive-definite matrix $\bm{L}$ is required to satisfy (\ref{eq:Lips}). From (\ref{eq:fq}), we have:
\begin{equation}\label{eq:gradientf}
    \nabla{f}(\bm{q}^g)=\bm{A}\Phi[\bm{A}\bm{q}^g+\bm{c}(\bm{d})-\bm{V}_r]
\end{equation}
It follows that:
\begin{equation}\label{eq:gradient}
     <\nabla{f}(\bm{q}^g)-\nabla{f}(\bm{y}),\bm{q}^g-\bm{y}>=||\bm{q}^g-\bm{y}||_{\bm{A}\Phi\bm{A}}^2
\end{equation}
From (\ref{eq:gradient}), it follows that (\ref{eq:Lips}) is satisfied if the condition {(\ref{eq:L1})} holds:
\begin{equation}\label{eq:L1}
    \bm{L}\succeq\bm{A}\Phi\bm{A}
\end{equation}
i.e., $\bm{L}-\bm{A}\Phi\bm{A}$ is semi-positive definite.

\noindent
\textbf{Proposition 3:} Suppose $F(\bm{q}^g)=f(\bm{q}^g)+g(\bm{q}^g)$ satisfies the following conditions:
\begin{itemize}
    \item \textbf{[P3.A]} [P2.A]-[P2.C] hold.
    \item \textbf{[P3.B]} $g(\bm{q}^g)$ is an indicator function, and for $\forall\bm{q}^g,\bm{y}\in\mathbb{R}^N$, there exists a positive definite matrix $\bm{H}$ satisfying:
    \begin{align}
            <\nabla{f}(\bm{q}^g)-\nabla{f}(\bm{y}),\bm{q}^g-\bm{y}>\geq||\bm{q}^g-\bm{y}||_{\bm{H}}^2
    \end{align}
\end{itemize}
Then the sequence $\{\bm{q}^g(k)\}$, generated by Algorithm 1: GFGM-Based VVC, satisfies:
\begin{equation}
||\bm{q}^g(k)-\bm{q}^{g\ast}||\leq\frac{2||\bm{q}^g(0)-\bm{q}^{g\ast}||_{\bm{L}}}{(k+1)\sqrt{\sigma_{\rm min}(\bm{H})}}
\end{equation}
where $\sigma_{\rm min}(\cdot)$ denotes the smallest eigenvalue.

\noindent
\textbf{Proof of Proposition 3:} See Appendix B.

With respect to Proposition 3, it shows $\bm{q}^g(k)$ will finally converge to the optimal solution $\bm{q}^{g\ast}$, indicating the system is stable when [P3.A] and [P3.B] hold. The condition [P3.A], i.e., [P2.A]-[P2.C], has been discussed in the previous analysis regarding Proposition 2. With respect to [P3.B], it follows from (\ref{eq:fq}) that:
\begin{equation}
    <\nabla{f}(\bm{q}^g)-\nabla{f}(\bm{y}),\bm{q}^g-\bm{y}>=||\bm{q}^g-\bm{y}||_{\bm{A}\Phi\bm{A}}^{2}
\end{equation}
It is clear that [P3.B] always holds as we set $\bm{H}=\bm{A}\Phi\bm{A}$.

{\textbf{In short, with respect to the VVC problem (\ref{eq:ProblemReform}), we can conclude that as long as $\bm{L}\succeq\bm{A}\Phi\bm{A}$ holds, Propositions 2 and 3 will hold.}}

Note that the linearized distribution power flow model is leveraged to convexify the optimal power flow problem and facilitate the  algorithm  design  and  theoretical  analysis. \textit{In Proposition 4, we further analyze the overall performance of Algorithm 1: GFGM-Based VVC on the actual nonlinear power flow.}
\\
\noindent
\textbf{Proposition 4:} Let ${\bm{\hat{q}}^{g\ast}}$, $m(\bm{\hat{q}}^{g\ast})$ be the optimal solution and value of problem (\ref{eq:NonlinearlVoltageProb}), and ${\bm{{q}}^{g\ast}}$, $f(\bm{{q}}^{g\ast})$ be the optimal solution and value of problem (\ref{eq:ProblemReform}). Assume the following conditions hold:
\begin{itemize}
    \item \textbf{[P4.A]} The error between the linearized power flow model and the exact nonlinear power flow model is bounded. That is, there exists a $\delta<\infty$ satisfying
    \begin{align*}
        ||h(\bm{q}^g,\bm{d})-h_l(\bm{q}^g,\bm{d})||_2\leq{\delta}, \text{where~}\bm{\underline{q}}^g\leq\bm{q}^g\leq\bm{\overline{q}}^g
    \end{align*}
    \item \textbf{[P4.B]} The error between the optimal objective values of problem (\ref{eq:NonlinearlVoltageProb}) and problem (\ref{eq:ProblemReform}) is bounded. That is, there exists a $\tau<\infty$ satisfying
    \begin{align*}
        \big|m({\bm{\hat{q}}^{g\ast}})-f({\bm{{q}}^{g\ast}})\big|\leq{\tau}
    \end{align*}
    \item \textbf{[P4.C]} [P2.A]-[P2.C] hold.
\end{itemize}
Then, it follows that:
\begin{equation}\label{eq:Stability}
    {m}(\bm{q}^g(k))-m(\bm{\hat{q}}^{g\ast})\leq\frac{1}{2}||\bm{E}||_2^2\delta^2+\frac{2||\bm{q}^g(0)-\bm{q}^{g\ast}||_{\bm{L}}^2}{(k+1)^2}+\tau
\end{equation}
where $\bm{E}$, satisfying $\bm{E}^T\bm{E}=\Phi$, is an upper triangular matrix with real and positive diagonal entries.
\\
\noindent
\textbf{Proof of Proposition 4:} See Appendix C.

In Proposition 4, $m(\bm{q}^g(k))$ can be regarded as the  objective value in the actual nonlinear power flow system after implementing $\bm{q}^g(k)$, which is determined based on the linearized power flow. Proposition 4 shows that the gap between $m(\bm{q}^{g}(k))$ and $m(\bm{\hat{q}}^{g\ast})$ is always bounded by three terms: (i) The error $\tau$ between the optimal values of problem (\ref{eq:NonlinearlVoltageProb}) considering the linearized power flow constraints, and problem (\ref{eq:ProblemReform}) considering the nonlinear power flow constraints; (ii) $\frac{1}{2}||E||_2^2\delta^2$ is in proportion to $\delta^2$; (iii) $\frac{2||\bm{q}^g(0)-\bm{q}^{g\ast}||_{\bm{L}}^2}{(k+1)^2}$ decreases as $k$ increases.

\section{Automatic Self-Adaptive Local Voltage Control Design}\label{sec:ALVC}
\subsection{Overview}
In this section, we mainly focus on the transition from Algorithm 1: GFGM-Based VVC to the self-adaptive local voltage control design. Note that $\gamma$ in $\textbf{S2}$ of Algorithm 1: GFGM-Based VVC can be simultaneously updated by each bus agent. And \textbf{S3} in Algorithm 1: GFGM-Based VVC is naturally decomposable, which can be locally updated by each bus agent $i$ in the way:
\begin{equation}
    {y}_i(k+1)={q}^g_i(k)+\big[\frac{\gamma(k)-1}{\gamma(k+1)}\big]\big[{q}_i^g(k)-{q}_i^g(k-1)\big], \forall{i}\in\mathcal{N}
\end{equation}
However, \textbf{S1} in Algorithm 1: GFGM-Based VVC is not naturally decomposable. The key challenge for local voltage control is how to design $\Phi$ and $\bm{L}$ such that \textbf{S1} in Algorithm 1: GFGM-Based VVC can also be locally implemented by each bus agent.
\subsection{Selection of $\Phi$ and \textbf{L}}
Regarding the choice of $\Phi$ and $\bm{L}$, there are two main considerations: (i) $\Phi$ and $\bm{L}$ should satisfy (\ref{eq:L1}) to make [P2.C] hold, thus ensuring the stability, convergence and optimality properties of voltage control; (ii) Under the selected $\Phi$ and $\bm{L}$, \textbf{S1} in Algorithm 1: GFGM-Based VVC can be locally implemented.

To this end, we first design $\bm{L}$ as a diagonal positive definite matrix, i.e., $\bm{L}=\text{diag}(L_{1},...,L_{N})$. As shown in Proposition 5, the diagonal positive definite matrix  $\bm{L}$ contributes to the local implementation of $\bm{q}^{g}(k)=p_{\bm{L}}(\bm{y}(k))$ in \textbf{S1} of Algorithm 1: GFGM-Based VVC. 

\noindent
\textbf{Proposition 5:} As $\bm{L}$ is a diagonal positive definite matrix, $\bm{q}^{g}(k)=p_{\bm{L}}(\bm{y}(k))$ in \textbf{S1} of Algorithm 1: GFGM-Based VVC is equivalent to:
\begin{equation}\label{eq:LocalImple}
    q_i^g(k)=\big[y_i(k)-\frac{1}{L_i}\frac{\partial{f}(\bm{y}(k))}{\partial{y}_i(k)}\big]_{\underline{q}_i(k)}^{\overline{q}_i(k)}, \forall{i}\in\mathcal{N}
\end{equation}
which can be expressed in a compact form:
\begin{equation}\label{eq:RewrittenS1}
    \bm{q}^g(k)=[\bm{y}(k)-\bm{L}^{-1}\nabla{f}(\bm{y}(k))]_{\bm{\underline{q}}^g}^{\bm{\overline{q}}^g}
\end{equation}

\noindent
\textbf{Proof of Proposition 5:} See Appendix D.

With respect to (\ref{eq:LocalImple}), the remaining challenge for the local implementation of $\bm{q}^{g}(k)=p_{\bm{L}}(\bm{y}(k))$ is locally calculating $\frac{\partial{f}(\bm{y}(k))}{\partial{y}_i(k)}$ when $\bm{L}$ is a diagonal positive definite matrix. 

Next, we further discuss how to choose $\bm{L}$ and $\Phi$ to resolve the dilemma: the local implementation of calculating $\frac{\partial{f}(\bm{y}(k))}{\partial{y}_i(k)}$. From (\ref{eq:gradient}), it follows that:
\begin{equation}\label{eq:fy}
    \nabla{f}(\bm{y}(k))=\bm{A}\Phi[\bm{A}\bm{y}(k)+\bm{c}-\bm{V}_r]
\end{equation}
It is obtained from Algorithm 1: GFGM-Based VVC that:
\begin{equation}\label{eq:yk}\small
    \bm{y}(k)=
    \begin{cases}
    &\bm{q}^g(0)=\bm{0}, k=1\\
    &\bm{q}^g(k-1)+\\&\big[\frac{\gamma(k-1)-1}{\gamma(k)}\big]\big[\bm{q}^g(k-1)-\bm{q}^g(k-2)\big], k\geq{2}
    \end{cases}
\end{equation}
Substituting (\ref{eq:yk}) into $\bm{A}\bm{y}(k)+\bm{c}-\bm{V}_r$:
\begin{subequations}\small
\begin{align}
    &\bm{A}\bm{y}(1)+\bm{c}-\bm{V}_r=\bm{A}\bm{q}^g(0)+\bm{c}-\bm{V}_r=\bm{V}(0)-\bm{V}_r\\
    \nonumber&\bm{A}\bm{y}(k)+\bm{c}-\bm{V}_r=\big[1+\frac{\gamma(k-1)-1}{\gamma(k)}\big][\bm{A}\bm{q}^g(k-1)+\bm{c}-\bm{V}_r]\\
    &\nonumber-\frac{\gamma(k-1)-1}{\gamma(k)}[\bm{A}\bm{q}^g(k-2)+\bm{c}-\bm{V}_r]\\
    &\nonumber=\big[1+\frac{\gamma(k-1)-1}{\gamma(k)}\big][\bm{V}(k-1)-\bm{V}_r]\\&-\frac{\gamma(k-1)-1}{\gamma(k)}[\bm{V}(k-2)-\bm{V}_r], k\geq{2}
\end{align}
\end{subequations}
As we set $\Phi=\bm{A}^{-1}$, (\ref{eq:fy}) can be expressed as follows:
\begin{subequations}\label{eq:fy2}\small
\begin{align}
    \nabla{f}(\bm{y}(1))&=\bm{V}(0)-\bm{V}_r\\
    \nonumber\nabla{f}(\bm{y}(k))&=\big[1+\frac{\gamma(k-1)-1}{\gamma(k)}\big][\bm{V}(k-1)-\bm{V}_r]\\&-\frac{\gamma(k-1)-1}{\gamma(k)}[\bm{V}(k-2)-\bm{V}_r], k\geq{2}
\end{align}
\end{subequations}
Thus, for $\forall{i}\in\mathcal{N}$, $\frac{\partial{f}(\bm{y}(k))}{\partial{y}_i(k)}$ can be calculated locally by:
\begin{subequations}\label{eq:localgradient}
\begin{align}
    \frac{\partial{f}(\bm{y}(1))}{\partial{y}_i(1)}&=V_i(0)-V_r\\
    \nonumber\frac{\partial{f}(\bm{y}(k))}{\partial{y}_i(k)}&=\big[1+\frac{\gamma(k-1)-1}{\gamma(k)}\big][{V}_i(k-1)-{V}_r]\\&-\frac{\gamma(k-1)-1}{\gamma(k)}[{V}_i(k-2)-{V}_r], k\geq{2}
\end{align}
\end{subequations}
It is clear that  $\frac{\partial{f}(\bm{y}(k))}{\partial{y}_i(k)}$ can be locally updated by each bus $i$ in (\ref{eq:localgradient}).

From the above analysis, we can conclude that \textbf{S1} in Algorithm 1: GFGM-Based VVC can be locally implemented with a diagonal positive definite matrix $\bm{L}$ and $\Phi=\bm{A}^{-1}$. Moreover, with respect to the choice of $\bm{L}$, $\Phi$ and $\bm{L}$ should satisfy (\ref{eq:L1}), thus we have:
\begin{equation}\label{eq:L2}
    \bm{L}\succeq\bm{A}\bm{A}^{-1}\bm{A}=\bm{A}
\end{equation}
$\bm{L}=\bm{A}$ provides the tightest bound $Q_{\bm{L}}(\bm{q}^g,\bm{y})$ for $F(\bm{q}^g)$, the best convergence performance of GFGM can be expected. However, such a selection cannot facilitate the local implementation of \textbf{S1} in Algorithm 1: GFGM-Based VVC. {Instead} $\bm{L}$ should be a diagonal positive definite matrix satisfying $\bm{L}\succeq\bm{A}$. Consequently, we utilize the following convex semi-definite programming problem to determine $\bm{L}$:
\begin{subequations}\label{eq:SDP}
\begin{align}
    \min_{\bm{L}}& \text{~tr}{\bm{L}}=\sum_{i=1}^N{L_i}\\
    \text{s.t.~}& \bm{L}\succeq\bm{A}, \bm{L}=\text{diag}(L_{1},...,L_{N})
\end{align}
\end{subequations}
\medskip
\noindent
\textbf{Remark 2:} In a nutshell, we select $\Phi=\bm{A}^{-1}$ and $\bm{L}$, determined by (\ref{eq:SDP}). Such a choice not only satisfies (\ref{eq:L1}) to hold [P2.C], but also facilitates the local implementation of Algorithm 1 :GFGM-Based VVC.

\subsection{Reinterpretation of GFGM: Modified Droop Control}
For ease of expression and analysis, we introduce $\mu({k})$ and $\bm{b}(k)$ as follows:
\begin{subequations}\label{eq:Mu}
\begin{align}
    \mu({k})&=
    \begin{cases}
        0, k=1\\
        \frac{\gamma(k-1)-1}{\gamma(k)}, {k}\geq{2}
    \end{cases}\\
    \bm{b}(k)&=
    \begin{cases}
    \bm{q}^g(0), k=1
    \\
    [1+\mu(k)]\bm{q}^g(k-1)-\mu(k)\bm{q}^g(k-2)\\
    +\mu(k)\bm{L}^{-1}[\bm{V}(k-2)-\bm{V}_r], {k}\geq{2}
    \end{cases}
\end{align}
\end{subequations}
Substituting (\ref{eq:yk}) and (\ref{eq:fy2}) into (\ref{eq:RewrittenS1}), we have:
\begin{subequations}\label{eq:qg2}\small
\begin{align}
    \nonumber\bm{q}^g(1)&=\big[-\bm{L}^{-1}[\bm{V}(0)-\bm{V}_r]+\bm{q}^g(0)\big]_{\bm{\underline{q}}^g}^{\bm{\overline{q}}^g}\\
    &=\big[-[1+\mu(1)]\bm{L}^{-1}[\bm{V}(0)-\bm{V}_r]+\bm{b}(1)\big]_{\bm{\underline{q}}^g}^{\bm{\overline{q}}^g}\\
    \nonumber\bm{q}^g(k)&=\big[-[1+\mu(k)]\bm{L}^{-1}[\bm{V}(k-1)-\bm{V}_r]\\
    &\nonumber+\bm{y}(k)+\mu(k)\bm{L}^{-1}[\bm{V}(k-2)-\bm{V}_r]\big]_{\bm{\underline{q}}^g}^{\bm{\overline{q}}^g}\\
    &=\big[-[1+\mu(k)]\bm{L}^{-1}[\bm{V}(k-1)-\bm{V}_r]+\bm{b}(k)\big]_{\bm{\underline{q}}^g}^{\bm{\overline{q}}^g}, k\geq{2}
\end{align}
\end{subequations}
For $\forall{i}\in\mathcal{N}$, it follows from (\ref{eq:qg2}) that: 
\begin{equation}\label{eq:k>=1}
      q_i^g(k)=\big[-a_i(k)[V_i(k-1)-V_r]+b_i(k)\big]_{\underline{q}_i^g}^{\overline{q}_i^g}, k\geq{1}
\end{equation}
with
\begin{subequations}\label{eq:abk>=1}
\begin{align}
  a_i(k)&=\frac{1+\mu(k)}{L_i}, k\geq{1}\\
  b_{i}(k)&=
  \begin{cases}
  {q}_i^g(0), k=1
  \\
  [1+\mu(k)]{q}_i^g(k-1)-\mu(k){q}_i^g(k-2)\\
  +\frac{\mu(k)}{L_i}[V_i(k-2)-{V}_r], {k}\geq{2}
  \end{cases}
\end{align}
\end{subequations}
Note that \emph{both $a_i(k)$ and $b_i(k)$ are updated locally by bus $i$ without the need for communication, only relying on the previous VAr outputs and voltage measurements of bus $i$}. In this case, \emph{Algorithm 1: GFGM-Based VVC} is equivalent to \emph{Algorithm 2: Automatic Self-Adaptive Local Voltage Control (ASALVC): Offline Implementation}.

\begin{figure}[t]
   \centering
   \includegraphics[width=0.8\columnwidth]{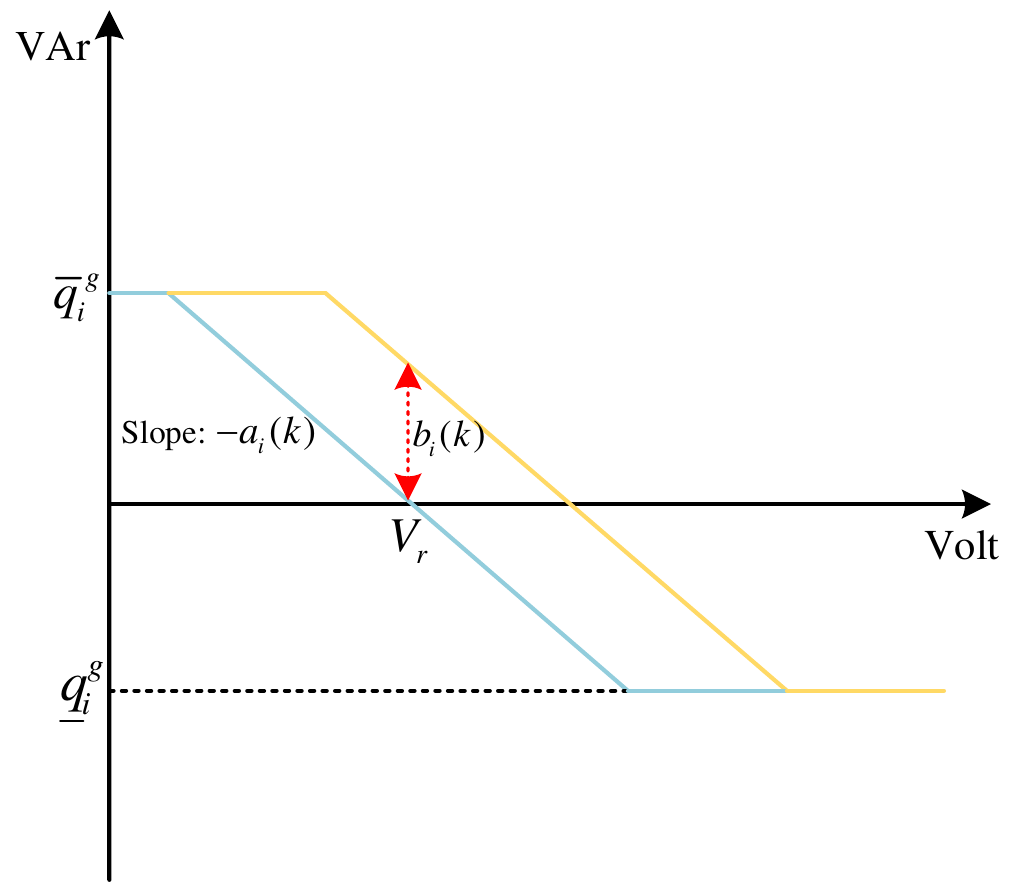}
   \caption{{Modified droop control with bus-specific time-varying coefficients}}
   \label{fig:DroopControl}
                          \end{figure}

\medskip
\noindent
\textbf{Remark 3:} Interestingly, this local voltage described in Algorithm 2 can be regarded as \textit{a modified voltage droop control with bus-specific self-adaptive coefficients}. As shown in Fig.\ref{fig:DroopControl}, the modified droop control (the yellow line segments) for bus $i$ is translated from the blue line segments with the slope $-a_i(k)$. 

\medskip
\noindent
\textbf{{Remark 4:}} {The proposed ASALVC is equivalent to the GFGM-Based VVC with  $\Phi=\bm{A}^{-1}$ and $\bm{L}$, determined by (\ref{eq:SDP}). As we discussed before, $\Phi=\bm{A}^{-1}$ and $\bm{L}$, determined by (\ref{eq:SDP}), can always satisfy all the conditions in Propositions 1-4, thus Propositions 1-4 always hold for the ASALVC by setting $\Phi=\bm{A}^{-1}$ and $\bm{L}$, determined by (\ref{eq:SDP}).}

\begin{algorithm}[t]
\renewcommand{\thealgorithm}{2}\selectfont
\small
\caption{Automatic Self-Adaptive Local Voltage Control (ASALVC): Offline Implementation}
\begin{algorithmic}
\STATE \hspace{-3mm}{\bf Initialization:} Set the iteration time $k=0$. Each bus $i$ sets
$\gamma(1)=1$, ${q}_i^g(0)={y}_i(1)={0}$,
\STATE \hspace{-3mm}{\bf For} {$k\geq1$}: Each bus $i$ alternately update variables by the following steps until convergence: 
\STATE\hspace{-1mm} {$\bullet$} Update $\mu(k)$, $a_i(k)$, $b_i(k)$ by (\ref{eq:Mu}a), (\ref{eq:abk>=1}a), (\ref{eq:abk>=1}b), respectively.
\STATE\STATE\hspace{-1mm} {$\bullet$} Update $q_i^g(k)$ by (\ref{eq:k>=1}), based on $V_i(k-1)$.
\STATE\STATE\hspace{-1mm} {$\bullet$} Update $\gamma(k+1)$:
\begin{align*}
    \gamma(k+1)=\frac{1+\sqrt{1+4\gamma(k)^2}}{2}
\end{align*}
\end{algorithmic}
\end{algorithm}

\begin{algorithm}[t]
\renewcommand{\thealgorithm}{3}\selectfont
\small
\caption{Automatic Self-Adaptive Local Voltage Control (ASALVC): Online Implementation}
\begin{algorithmic}
\STATE \hspace{-3mm}{\bf For} any bus $i$ at time step $t$:
\STATE\hspace{-1mm} {$\bullet$} Estimate VAr Limits: Locally update $\underline{q}^g_i$ and $\overline{q}^g_i$
\STATE\hspace{-1mm} {$\bullet$} Reset $\gamma(t)$: If $t$ $mod$ $T_\gamma$ =0, then set $\gamma(t)=1$.
\STATE\hspace{-1mm} {$\bullet$} Reset $\mu(t)$: If $t$ $mod$ $T_\gamma$ =0, then set $\mu(t)=0$; otherwise update $\mu(t)$ by $\frac{\gamma(t-1)-1}{\gamma(t)}$.
\STATE\hspace{0mm}{$\bullet$} Update  $a_i(t)$, $b_i(t)$ by (\ref{eq:abk>=1}a), (\ref{eq:abk>=1}b), respectively. 
\STATE\hspace{-1mm} {$\bullet$} Update $q_i^g(t)$ by (\ref{eq:k>=1}), based on the voltage measurement $V_i(t-1)$ at time $t-1$.
\STATE\hspace{-1mm} {$\bullet$} Update $\gamma(t+1)$:
\begin{align*}
    \gamma(t+1)=\frac{1+\sqrt{1+4\gamma(t)^2}}{2}
\end{align*}
\end{algorithmic}
\end{algorithm}
                          
                      \begin{figure}[htb]
   \centering
   \includegraphics[width=1\columnwidth]{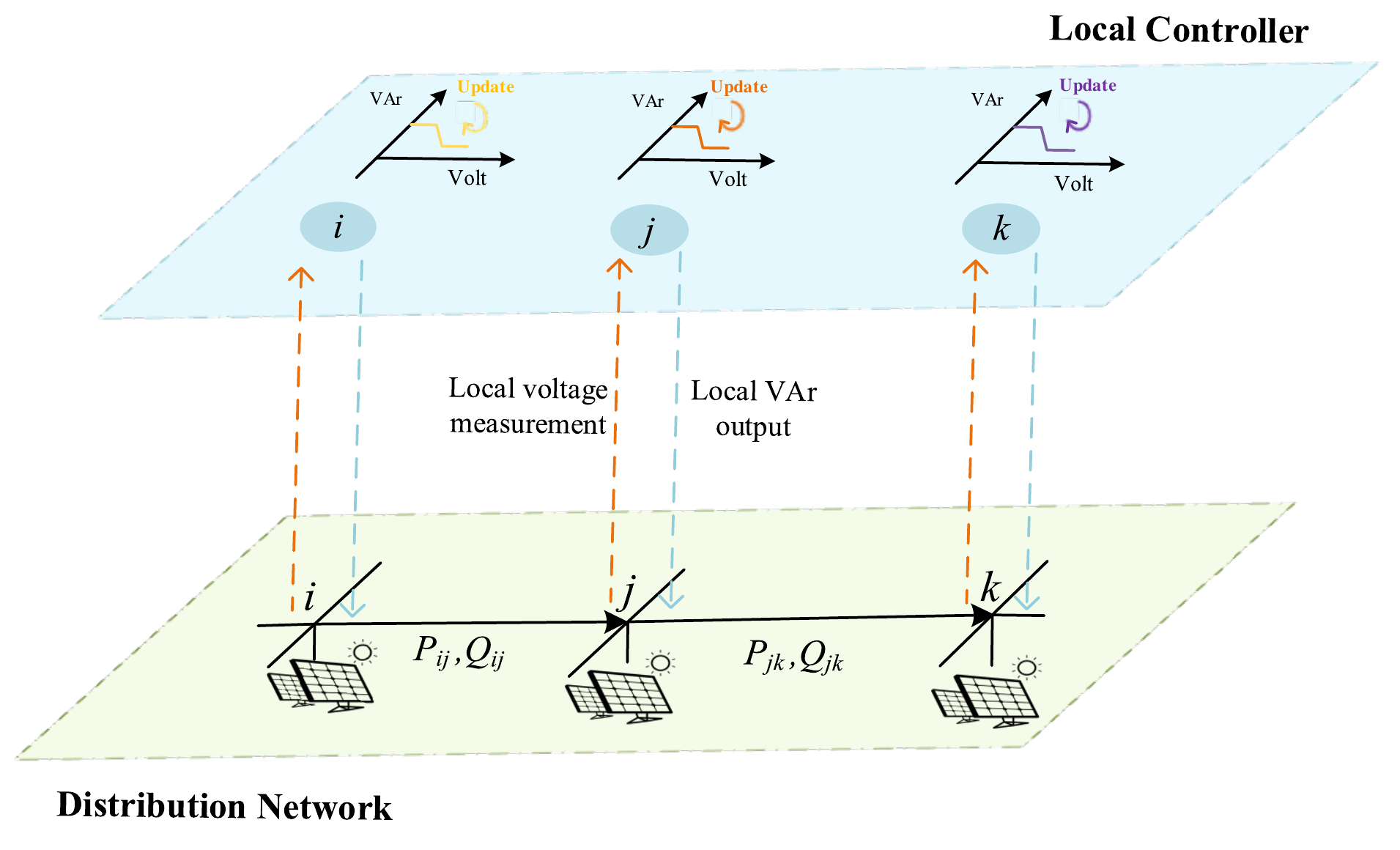}
   \caption{{Automatic Self-Adaptive Local Voltage Control Demonstration.}}
   \label{fig:Controller}
                          \end{figure}
                          
\subsection{Online Implementation}
To better deal with the time-varying system variations, associated with the time-varying $\bm{d}$, the online implementation of Algorithm 2 is proposed in this subsection.

\medskip
\noindent
{\textbf{Remark 5:} With respect to offline implementation, the decision/control variables are not applied to the physical world until those variables converge \cite{OV2}. However, with respect to online implementation, the decision/control variables are adjusted in real-time (for each iteration), based on the real-time feedback from operating statuses, to adapt to real-time changes in the environment.
}

The online implementation of ASALVC consists of the following key steps:

\textit{1)} ${\textit{Estimate } {\underline{q}}^g_i \textit{ and } {\overline{q}}^g_i}:$
For the online implementation, VAr limits $\underline{q}^g_i$ and $\overline{q}^g_i$ are updated based on the inverter capacities and the instantaneous real power outputs of DERs. This online update of VAr limits ensures to satisfy the inverter capacity limit in a time-varying system, preventing inverters from overloading.

\textit{2)} {\textit{Reset} $\gamma$ and $\mu$:}
With respect to the offline implementation, we usually start with $\gamma=1$, $\mu=0$, and then $\gamma$ is updated in each iteration. For the online implementation, we propose to reset  $\gamma=1$ and $\mu=0$ every $T_{\gamma}$ time steps to ensure the tracking capability.

\textit{3)} {\textit{Update} $a_i$, $b_i$, $q_i^g$ and $\gamma$:} 
Update $a_i$, $b_i$ by (\ref{eq:abk>=1}), $q_i^g$ by (\ref{eq:k>=1}), $\gamma$ by $\gamma(t+1)=\frac{1+\sqrt{1+4\gamma(t)^2}}{2}$. With respect to the online implementation, the actual voltage measurement $V_i$, instead of the one calculated by the linearized distribution power flow, is used to update $b_i$ and $q_i^g$ in (\ref{eq:k>=1})-(\ref{eq:abk>=1}).

 \begin{table*}[htb]
\normalsize
\centering
\caption{Droop Control Comparisons}\label{Table:Comparisons}
\footnotesize\small
\renewcommand\arraystretch{1.0}
        \begin{tabular}{cccc}
        \specialrule{0.05em}{0mm}{0.5mm}
        \specialrule{0.05em}{0.5mm}{1mm}
         Control Type&Update&Description&Optimality\\
          \specialrule{0.05em}{1mm}{1mm}
        CDC \cite{AdvancedVVC}-\cite{1547-2018}&$q_i(t+1)=\big[-a_i[V_i(t)-V_r]\big]_{\underline{q}_i^g}^{\overline{q}_i^g}$& Constant slope, constant intercept& w/o  analyses\\
        \specialrule{0.05em}{1mm}{1mm}
        DDC \cite{PJ}&$q_i(t+1)=(1-\alpha_i)q_i(t)+\alpha_i\big[-a_i[V_i(t)-V_r]\big]_{\underline{q}_i^g}^{\overline{q}_i^g}$&Constant slope, time-varying intercept& w/o analyses\\
        \specialrule{0.05em}{1mm}{1mm}
         GPDC\cite{YG,HZ}&$q_i(t+1)=\big[q_i(t)-a_i[V_i(t)-V_r]\big]_{\underline{q}_i^g}^{\overline{q}_i^g}$&Constant slope, time-varying intercept&w/  analyses\\
         \specialrule{0.05em}{1mm}{1mm}
         SGPDC\cite{HZ}&$q_i(t+1)=\big[q_i(t)-a_id_i[V_i(t)-V_r]\big]_{\underline{q}_i^g}^{\overline{q}_i^g}$&Constant slope, time-varying  intercept&w/ analyses\\
         \specialrule{0.05em}{1mm}{1mm}
         ASALVC&$q_i(t+1)=\big[-a_i(t)[V_i(t)-V_r]+b_i(t)\big]_{\underline{q}_i^g}^{\overline{q}_i^g}$&Time-varying slope, time-varying  intercept&w/ analyses\\
         \specialrule{0.05em}{0.75mm}{0.5mm}
         \specialrule{0.05em}{0.5mm}{0mm}
        \end{tabular}
\end{table*}

The details of the online implementation are provided in \emph{Algorithm 3: Automatic Self-Adaptive Local Voltage Control (ASALVC): Online Implementation} and Fig.\ref{fig:Controller}. As depicted in Fig.\ref{fig:Controller}, each bus agent locally measures its voltage, then updates coefficients, i.e., $a_i(t)$ and $b_i(t)$,of its modified voltage control based on its voltage measurement, {and} finally determines its local VAr output. {As shown in (\ref{eq:abk>=1}), $a_i(t)$ and $b_i(t)$ can be updated online by simple arithmetic based on the previous VAr outputs, $q_i(t-1)$ and $q_i(t-2)$, and the previous voltage measurement $V_i(t-2)$,
}

\medskip
\noindent
\textbf{Remark 6:} Though we design ASALVC and establish its theoretical analysis, based on the linearized distribution power flow, under a fixed condition, the online implementation could asymptotically mitigate the model errors due to the closed-loop nature. More specifically, the online implementation of ASALVC can track changing network conditions as these changes manifest themselves in the network state, i.e., the actual voltage measurement, that is used to compute the control solution, i.e., VAr outputs of DERs.

\subsection{Comparisons with other droop controls}
The common existing droop controls can be roughly classified into the following types:

(1) Classical Droop Control \cite{AdvancedVVC}-\cite{1547-2018} (CDC): As advocated in IEEE 1547-2018 Standard, the CDC adjusts the inverter VAr output based on the instantaneous bus voltage mismatch. The CDC with zero dead band is updated by $q_i(t+1)=\big[-a_i[V_i(t)-V_r]\big]_{\underline{q}_i^g}^{\overline{q}_i^g}$, associated with a constant slope and a constant intercept. But it cannot guarantee optimum and always suffers from stability problems.

(2) Delayed Droop Control \cite{PJ} (DDC): The DDC can address the instability issues of CDC to a great degree. The VAr output of DDC depends on a weighted combination of the previous voltage and VAr output: $q_i(t+1)=(1-\alpha_i)q_i(t)+\alpha_i\big[-a_i[V_i(t)-V_r]\big]_{\underline{q}_i^g}^{\overline{q}_i^g}$, associated with a constant slope and a time-varying intercept, where $0<\alpha_i<1$ is a weighted parameter. However, the work \cite{PJ} does not provide optimality analyses. 

(3) GP-Based Droop Control \cite{YG,HZ} (GPDC): The GP update is applied to the droop control, thus generating the GPDC. It is updated by $q_i(t+1)=\big[q_i(t)-a_i[V_i(t)-V_r]\big]_{\underline{q}_i^g}^{\overline{q}_i^g}$, associated with a constant slope and  a time-varying intercept. Optimality analyses are provided in \cite{YG,HZ}.

(4) Scaled GP-Based Droop Control \cite{HZ} (SGPDC): The scaled GP update is applied to the droop control to speed up the convergence rate of GPDC, facilitating the development of SGPDC. The inverse of the diagonals of Hessian matrix is always a popular choice to scale gradients. It is updated by $q_i(t+1)=\big[q_i(t)-a_id_i[V_i(t)-V_r]\big]_{\underline{q}_i^g}^{\overline{q}_i^g}$, associated with a constant slope and  a time-varying intercept, where $d_i$ is a scaled parameter. Optimality analyses are provided in \cite{HZ}.

Compared to those droop control methods, the ASALVC is associated with the time-varying slope $-a_i(t)$ and the time-varying intercept $b_i(t)$, significantly increasing the diversity and flexibility of local voltage control. In addition, as discussed in Proposition 2, the convergence rate of ASALVC is characterized by $O(1/k^2)$, it is faster than the GPDC and SGPDC, characterized by $O(1/k)$. The summary of droop control comparisons is provided in Table \ref{Table:Comparisons}.

\section{Case Study}
\label{sec:Case}
\subsection{Overview}
\begin{figure}[htb]
   \centering
   \includegraphics[width=1\columnwidth]{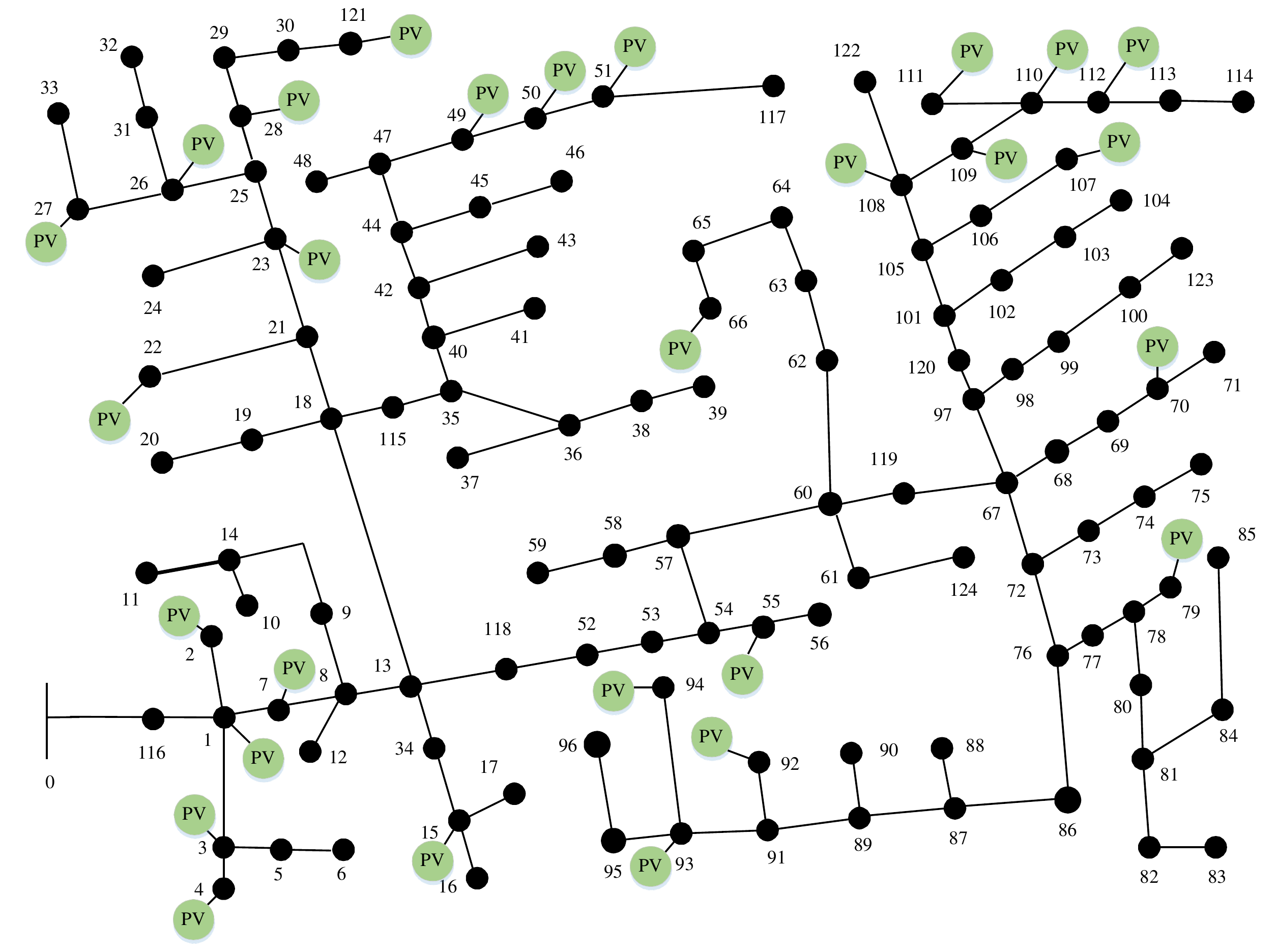}
   \caption{{Modified single-phase IEEE 123-bus test system.}}
   \label{fig:IEEE123}
                          \end{figure}

In this section, numerical simulations are performed in the modified single-phase IEEE 123-bus test system to validate the effectiveness and superiority of the proposed ASALVC. As shown in Fig.\ref{fig:IEEE123}, PV generators are distributed across the radial distribution network. 

In the numerical simulations, the base voltage for the network is $4.16$ kV and the base power is $100$ kVA. We set the reference of voltage magnitude as $\bm{V}_r=\bm{1}_{N}$, a $N\times1$ column vector of ones. Though the algorithm design of this paper is built on the linearized power flow model (\ref{eq:linear}), we simulate the ASALVC with the nonlinear AC power flow model (\ref{eq:nonlinear}) using MATPOWER \cite{MATPOWER}. Note that the actual bus voltage magnitude obtained from MATPOWER, instead of the one obtained from the linearized power flow model, is used as the voltage measurement to update the VAr outputs of DERs.
\subsection{Static Scenario}
In the static scenario, each bus has a constant load $1+j0.5$ kVA and each PV inverter can supply or absorb at most 10 kVAr. Different droop controls are considered for comparison, including the CDC, DDC, GPDC, SGPDC, ASALVC. {In addition, the centralized optimization is applied to directly solve this VVC problem (\ref{eq:ProblemReform}) through the CPLEX solver \cite{cplex}.
}
For the CDC and GPDC, we set $a_i=1$. For the DDC, we set $a_i=1$ and $\alpha_i=0.1$. For the SGPDC, we set $a_i=0.01$, $d_i=[\bm{A}\Phi\bm{A}]_{ii}^{-1}$, where $[\bm{A}\Phi\bm{A}]_{ii}$ is $i$-th row and $i$-th column element of $\bm{A}\Phi\bm{A}$, where the inverse of the diagonals of Hessian matrix is applied to scale gradients. With respect to the offline implementation of ASALVC in the static scenario, there is no need to manually set droop parameters as those parameters are automatically determined and adjusted by (\ref{eq:abk>=1}).

                      \begin{figure}[htb]
   \centering
   \includegraphics[width=0.9\columnwidth]{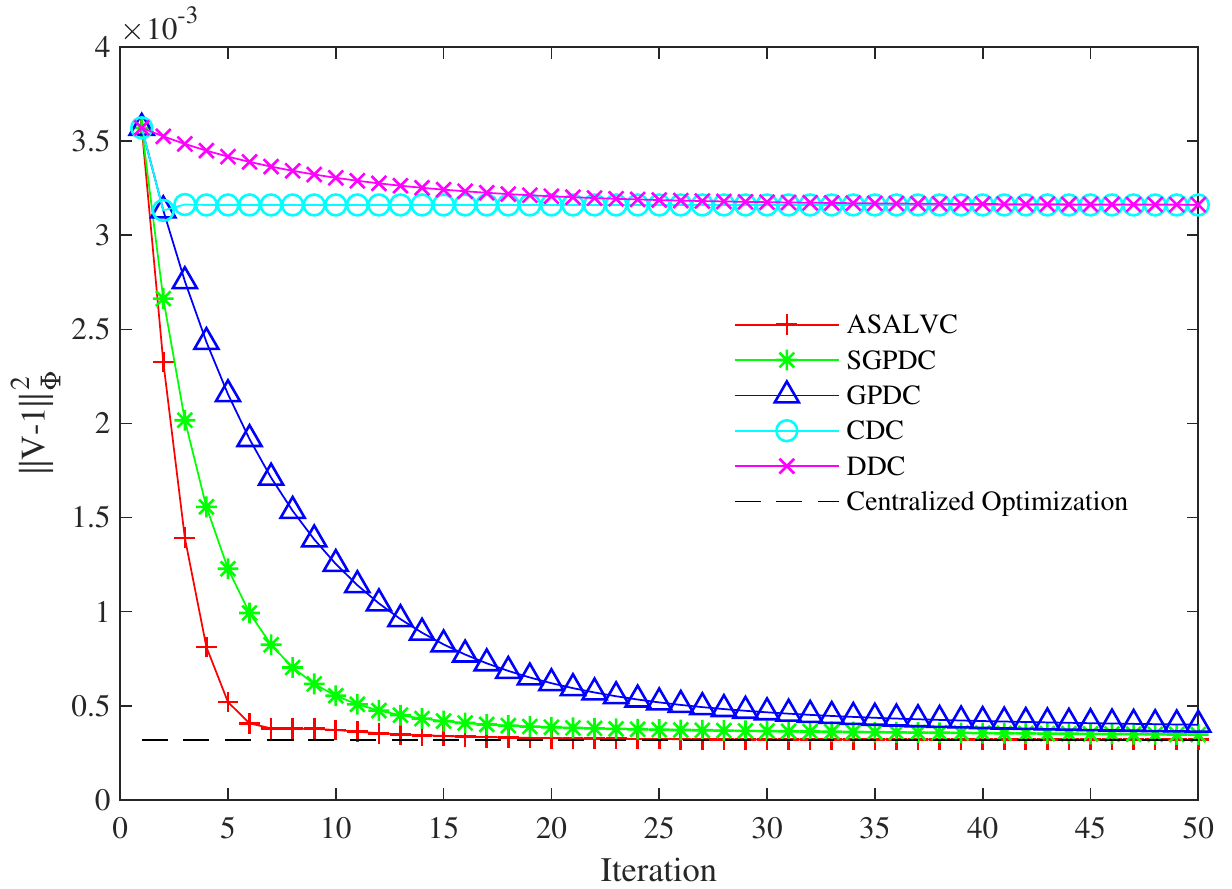}   \caption{{Voltage mismatch error versus iteration for various VAr controls under the static scenario.}}
   \label{fig:StaticComparison}
                          \end{figure}

As depicted in Fig.\ref{fig:StaticComparison}, the voltage mismatch error of CDC and DDC is around $3\times{10}^{-3}$. In contrast, the ASALVC, SGPDC, GPDC converge to voltage mismatch errors which are far less than the CDC and DDC.
{The convergence outcomes of ALALVC, SGPDC, and GPDC 
closely track the centralized optimization outcome, but the convergence outcomes of CDC and DDC do not. As shown in Table \ref{Table:Comparisons}, the design of CDC and DDC does not take into account the optimality property, thus the CDC and DDC do not show a good performance in terms of optimality.
} Besides, the ASALVC exhibits the best convergence performance
compared to other controls, reaching convergence after 6 iterations,
which is consistent with our previous theoretical analysis.
\subsection{Dynamic Scenario}
In the dynamic scenario, the time-varying system variations are considered. In this scenario, the capacities of PV inverters are set as 50 kVA, and  VAr limits $\bm{\underline{q}}^g,\bm{\overline{q}}^g$ are updated online based on the given inverter capacities and the instantaneous real power of PV generators. With respect to the online implementation of ASALVC in the dynamic scenario, we set $T_r=6s$, and the VAr outputs of DERs are updated every second.

\begin{figure}[t]
   \centering
   \includegraphics[width=0.9\columnwidth]{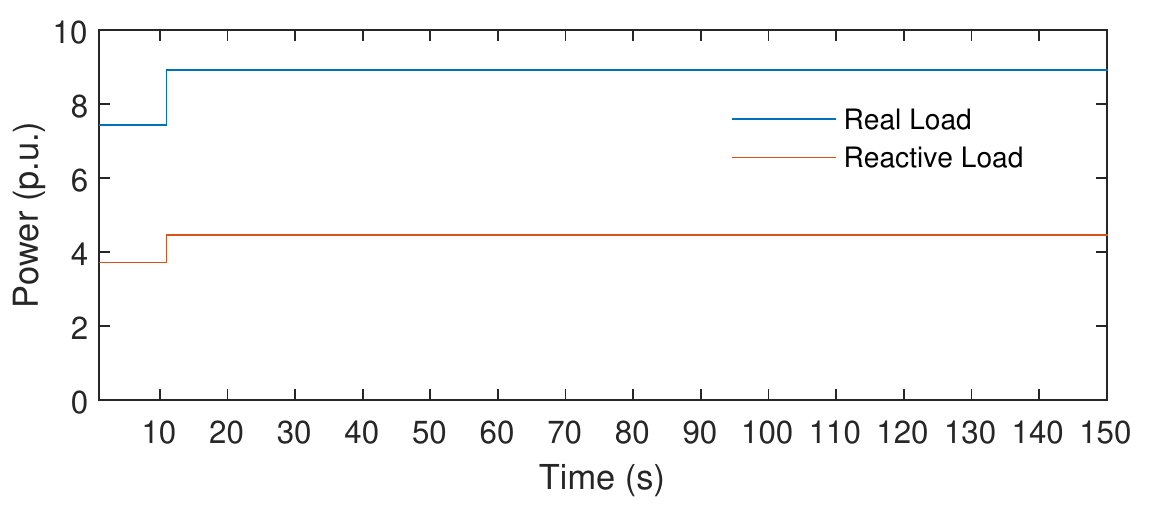}
   \caption{{Aggregate load in the dynamic scenario with a sudden load change.}}
   \label{fig:ChangeLoad}
                          \end{figure}
                       \begin{figure}[t]
   \centering
   \includegraphics[width=0.9\columnwidth]{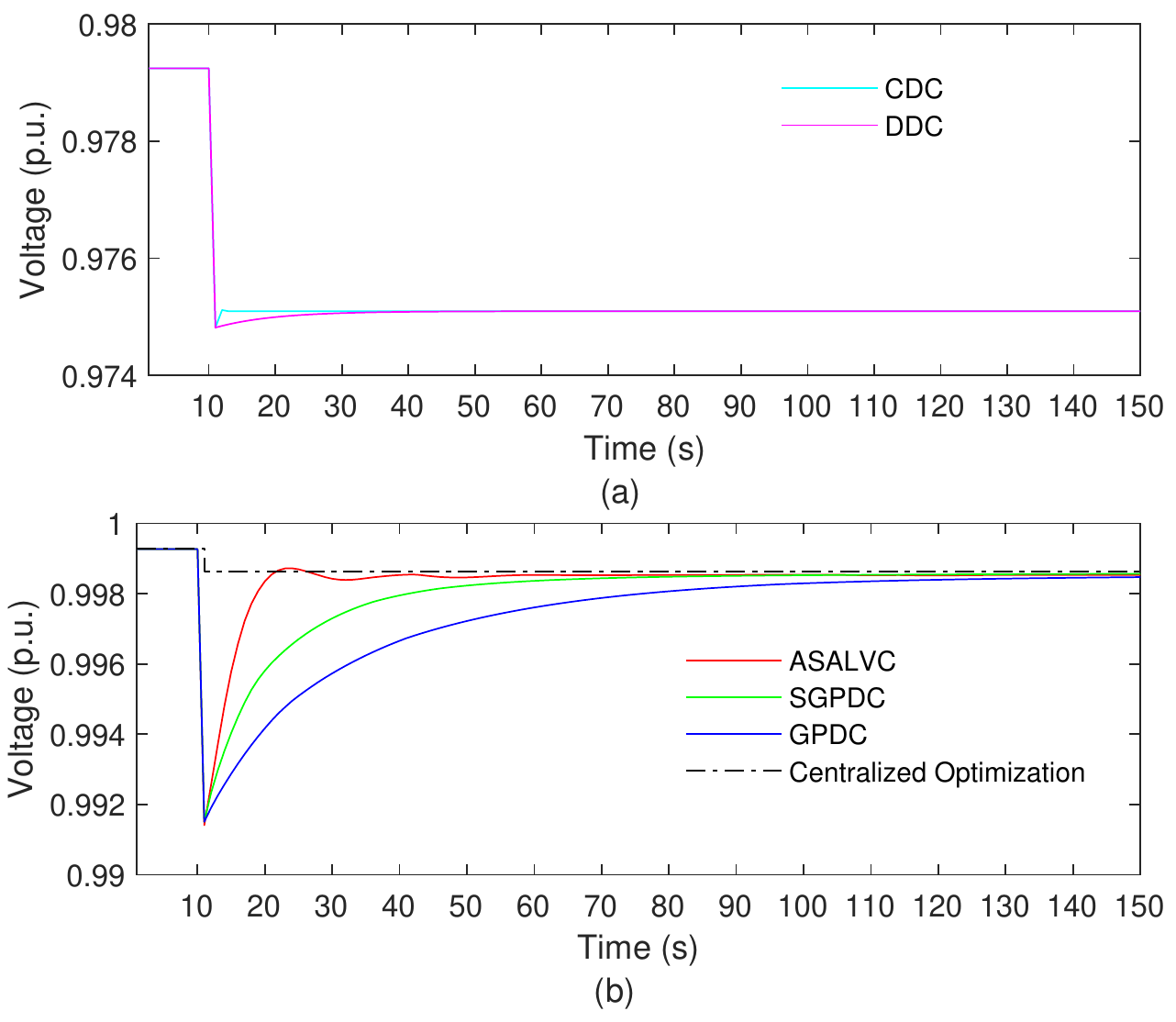}
   \caption{{The voltage at bus 56 under the dynamic scenario with a sudden load change: (a) for the CDC and DDC; (b) for the GPDC, SGPDC, ASALVC, and centralized optimization.}}
   \label{fig:V_Change}
                          \end{figure}

{First,  we consider a sudden load change in the modified IEEE 123-bus test system. As shown in Fig.\ref{fig:ChangeLoad}, suppose at $t=10 s$, the aggregate load suddenly increases. Taking the voltage at bus 56 as an example, we test the responses from different methods to this sudden load change. For those controls, the parameter settings follow from the static scenario and the VAr outputs of DERs are also updated every second. As shown in Fig.\ref{fig:V_Change}(a), the system is back to stable operations after some adjustments through the CDC and DDC as there is a sudden load change. However, the stable operating statuses, determined by the CDC and DDC, before and after this sudden load change are both far away from the optimal operating statuses, determined by the centralized optimization, before and after this sudden load change, where the optimal operating statuses are depicted as the black dotted line in Fig.\ref{fig:V_Change}(b). This is due to the fact the CDC and DDC can only guarantee the stability under some conditions but 
not optimality. On the contrary, as depicted in Fig.\ref{fig:V_Change}(b), the stable operating statuses, determined by the ASALVC, SGPDC, and GPDC, before and after this sudden load change are the same as the optimal operating statuses, determined by the centralized optimization, before and after this sudden load change. Besides, with the help of the ASALVC, the voltage at bus 56 around $t=20s$ closely tracks the optimal voltage at bus 56, determined by the centralized optimization, As there is a sudden load change, the ASALVC can be back to the optimal operating statuses with less time compared to the GPDC and SGPDC, indicating its stronger capability to quickly recover from such a sudden disturbance.
}

\begin{figure}[tb]
   \centering
   \includegraphics[width=0.9\columnwidth]{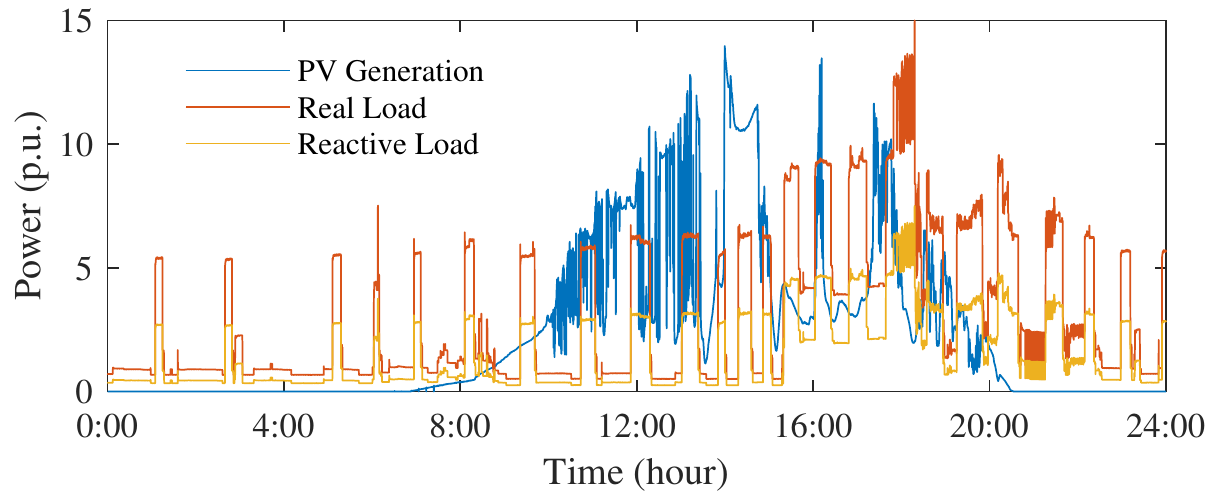}
   \caption{{Aggregate load and PV generation with continuous fast time-varying system changes.}}
   \label{fig:DynamicLoad}
                          \end{figure}

\begin{figure}[t]
   \centering
   \includegraphics[width=0.9\columnwidth]{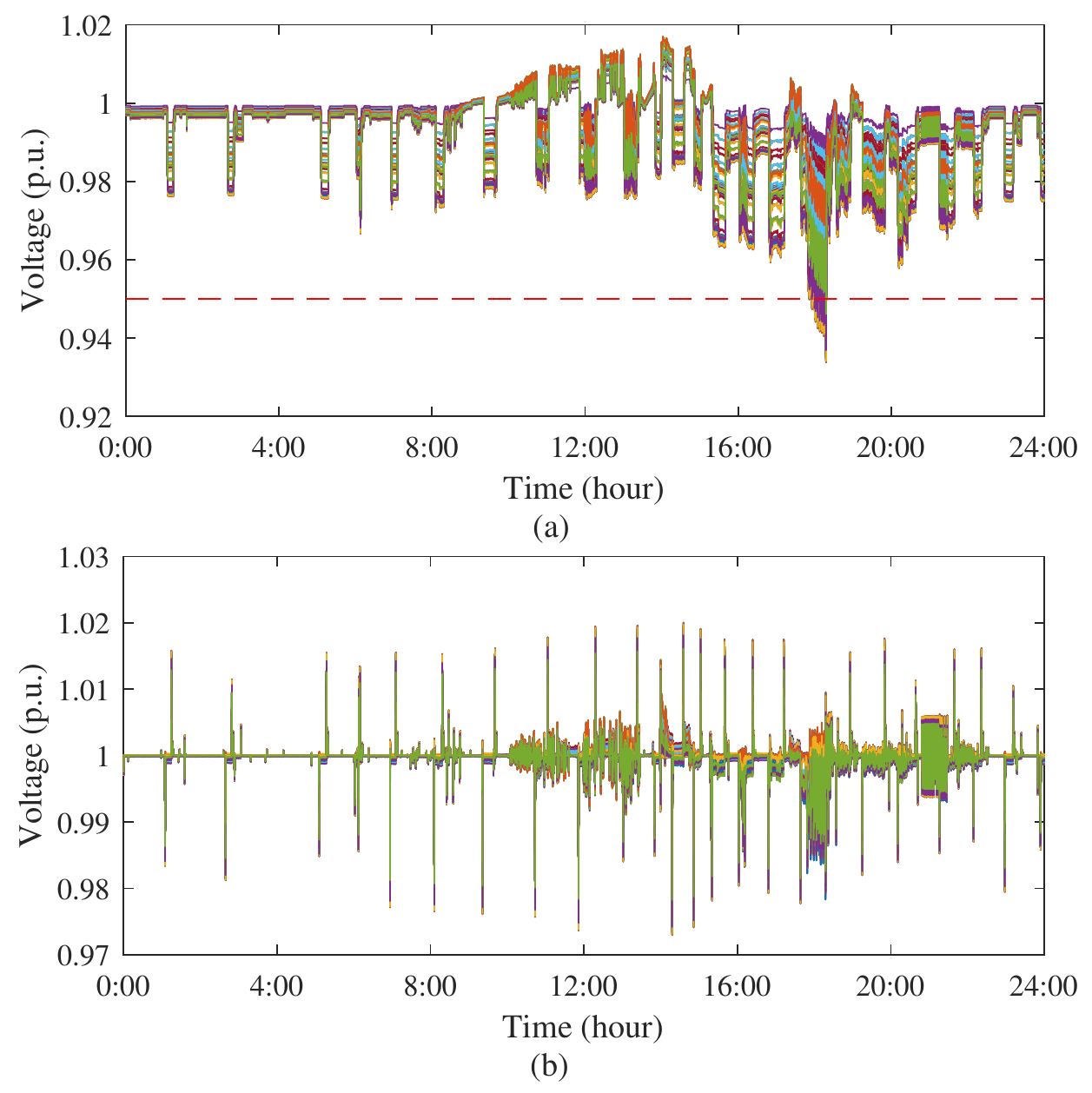}
   \caption{{The network voltage profile across the modified IEEE 123-bus test system (each curve depicts the voltage magnitude fluctuation for each  bus): {(a) Without any control; (b) With the ASALVC.}}
   }
   \label{fig:DynamicV}
                          \end{figure}

\begin{figure}[t]
   \centering
\includegraphics[width=0.9\columnwidth]{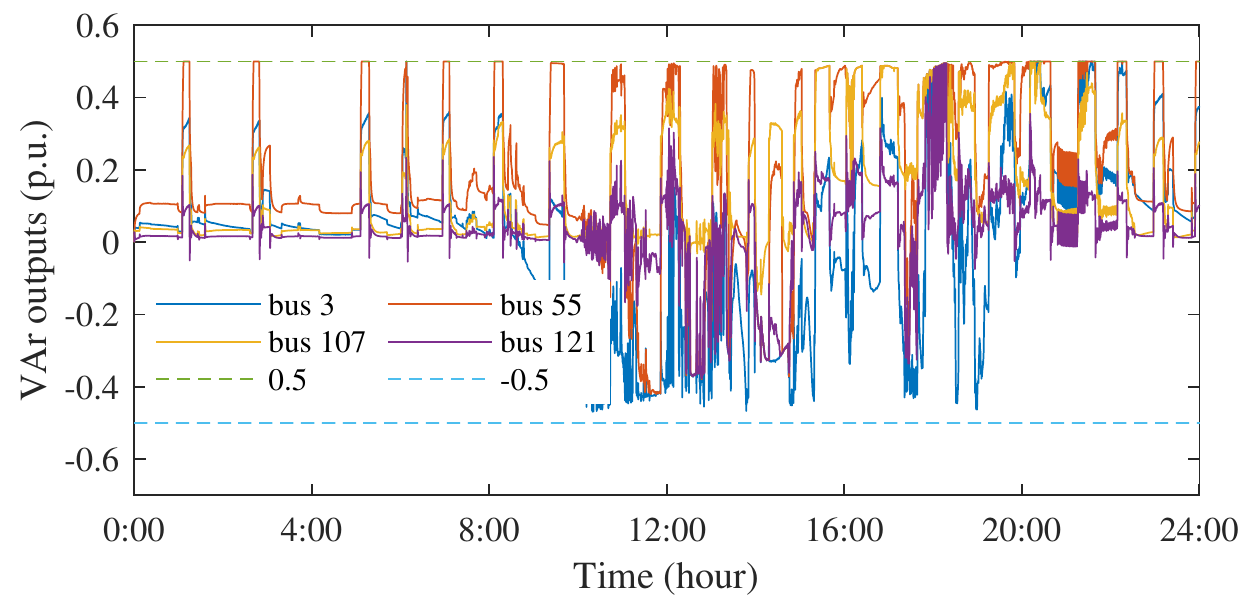}
   \caption{{VAr outputs of PV inverters at buses 3, 55, 107, 121 by using the ASALVC.
   }
   }
   \label{fig:DynamicQ}
                          \end{figure}

{Next, we consider a more realistic and complex system with continuous fast system changes.}  The aggregate load and PV generation with continuous fast system changes, distributed across the modified IEEE 123-bus test system,  are shown in Fig.\ref{fig:DynamicLoad}, where the time span is one day (24 hours) and the time granularity is 6 s. That is, the load and PV generation change rapidly every 6 seconds. Fig.\ref{fig:DynamicV} shows the network voltage profiles with the ASALVC and without any control. As seen in Fig.\ref{fig:DynamicV}(a), there are voltage violations for the test system without any control around 18:00. However, it is observed from Fig.\ref{fig:DynamicV}(b) and Fig.\ref{fig:DynamicQ}, despite the volatility in load and PV generation, the ASALVC can still effectively resolve voltage violation problems, not violating the capacity constraint.
                       \begin{figure}[t]
   \centering
\includegraphics[width=0.8\columnwidth]{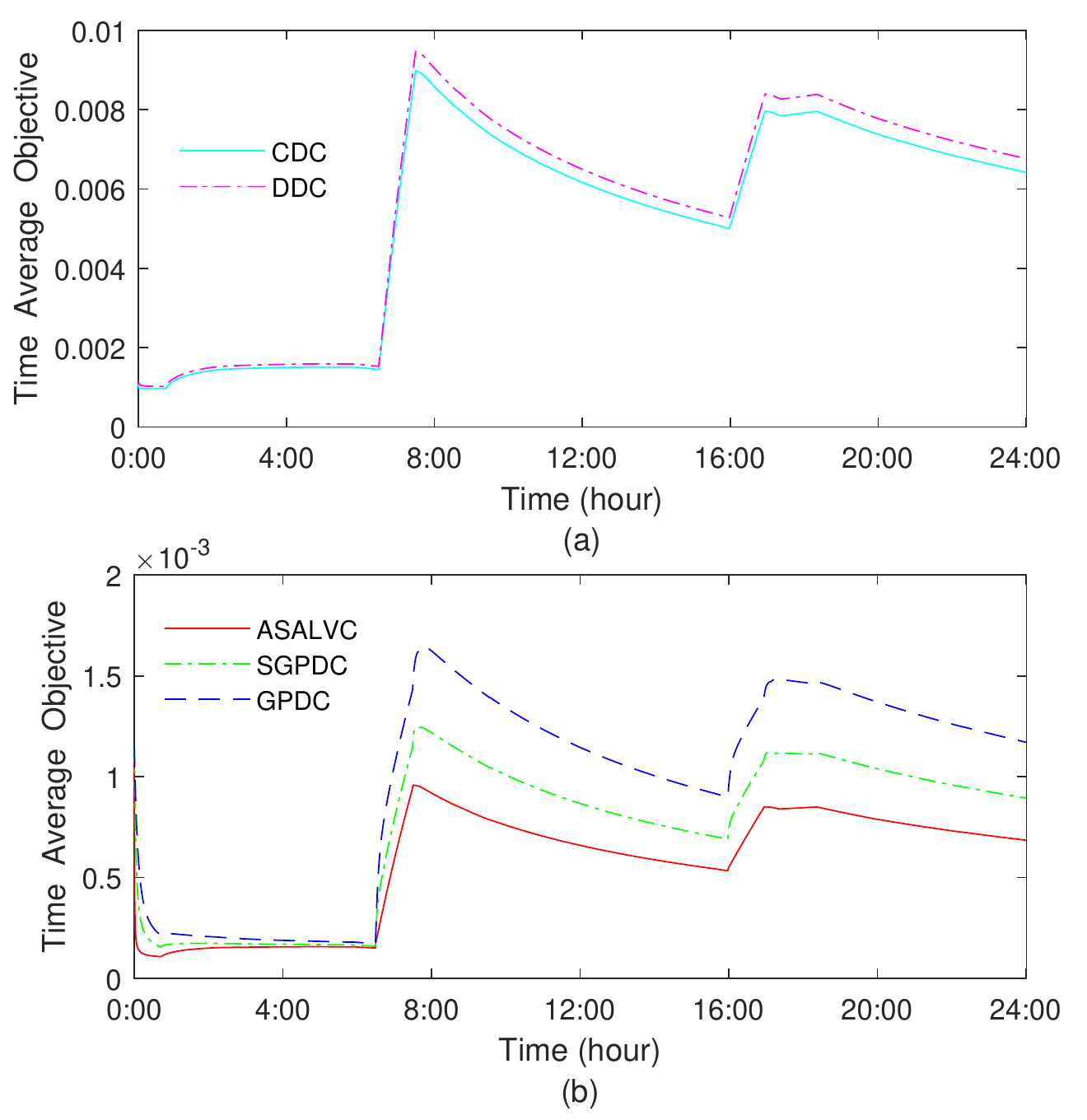}
   \caption{{Time average objectives for various VAr controls under the dynamic scenario with continuous fast system changes: (a) the performances of the CDC and DDC; (b) the performances of the GPDC, SGPDC, and ASALVC.}
   }
   \label{fig:DynamicComparison}
                          \end{figure}
\begin{table}[h]
\normalsize
\centering
\caption{{Voltage and capacity issues under the dynamic scenario with continuous fast system changes}}\label{Table:DynamicComparisons}
\footnotesize
\renewcommand\arraystretch{1.0}
        \begin{tabular}{cccccc}
        \hline
        \hline
          & CDC & DDC& GPDC&SGPDC&ASALVC \\
        \hline
        Voltage issue& Yes&Yes&No&No&No\\
        Capacity issue& No&No&No&No&No\\
        \hline
        \hline
        \end{tabular}
\end{table}

{For comparison, other droop controls, including the CDC and DDC, GPDC, SGPDC, are taken into account to handle continuous fast system variations.}\footnote{{Considering the centralized optimization is implemented after convergence, it is not suitable to apply the centralized optimization to the dynamic scenario with continuous fast system changes, thus the centralized optimization is not carried out here.}} {As shown in Table \ref{Table:DynamicComparisons}, the CDC and DDC suffer from voltage violation problems under the dynamic scenario with continuous fast system changes.
} {{For the CDC, DDC, GPDC, SGPDC, and ASALVC, there are not capacity violation problems in the online implementation as the VAr limits can be updated online based on the given inverter capacities and the instantaneous real power of PV generators.}} The time average objective \footnote{{The time average objective is the average objective value across time. The lower the time average objective is, the better its performance is.}}
across one day is selected as the metric to evaluate control performances.  {As shown in Fig.\ref{fig:DynamicComparison}, the performances of the CDC and DDC are poor under the dynamic scenario with continuous fast system changes. What explains this phenomenon? On the one hand, the CDC and DDC are associated with continuous slopes and intercepts, those droop control functions are not flexible to make full use of inverters' VAr outputs for voltage regulation. On the other hand, the CDC and DDC always suffer from stability and optimality problems, resulting in the poor performance in tracking the continuous fast changes. Compared to the CDC and DDC, the GPDC, SGPDC, ASALVC show better performances due to the capability of considering the optimality property.}  As shown in Fig.\ref{fig:DynamicComparison}, the ASALVC still exhibits the best performance compared to the GPDC and SGPDC, indicating it is more capable of maintaining a flat network voltage profile even for the dynamic scenario with continuous fast system changes. {As we discussed before, the ASALVC shows a faster convergence rate than other controls, leading to a greater tracking capability for continuous fast 
system changes in the dynamic scenario. In addition, the ASALVC is associated with both the time-varying slope and intercept, making itself more flexible to adapt to continuous fast system changes.
}


\section{Conclusion}
\label{sec:Conclusion}
This paper proposes an ASALVC strategy, where each bus agent locally adjusts the VAr output of its DER based on its time-varying voltage droop function. This voltage droop function is associated with the bus-specific time-varying slope and intercept, which can be dynamically updated merely based on the local voltage measurement. The dynamic adjustment characteristic  enables the ASALVC to track time-varying system changes.  Stability, convergence and optimality properties of this local voltage control are analytically established. 
Through numerical case studies, it shows the ASALVC always exhibits the best performance compared to other controls in both static and dynamic scenarios, validating its effectiveness and superiority. {Note that the proposed ASALVC can be further embedded into the two-layer VVC framework to consider discrete voltage regulation devices. In the upper layer, conventional discrete voltage regulation devices are scheduled  over a slow timescale. In the lower layer, the VAr outputs of DERs can be modulated by our proposed ASALVC, where the setting of discrete voltage regulation devices is maintained at the upper-layer solution. Our future research will focus on meshed distribution networks and the data-driven learning-assisted implementation of local voltage control.} 
\section*{Appendix A}
\noindent
\textit{Proof of Proposition 1:}

We introduce the function $z(\bm{x})=\frac{1}{2}\bm{x}^T\bm{L}\bm{x}-f(\bm{x})$. Since $f(\bm{x})$ is continuously differentiable, we can know $z(\bm{x})$ is continuously differentiable. By exploiting \cite[Theorem 2.1.3]{YN}, it follows that $z(\bm{x})$ is convex if and only if $<\nabla{z}(\bm{q}^g)-\nabla{z}(\bm{y}),\bm{q}^g-\bm{y}>\geq{0}$ holds for $\forall{\bm{q}^g,\bm{y}}\in\mathbb{R}^N$, which can be presented as follows:
\begin{equation}\label{eq:ProofProp1.2}
    \begin{split}
       &<\nabla{z}(\bm{q}^g)-\nabla{z}(\bm{y}),\bm{q}^g-\bm{y}>=\\&-<\nabla{f}(\bm{q}^g)-\nabla{f}(\bm{y}),\bm{q}^g-\bm{y}>+||\bm{q}^g-\bm{y}||_{\bm{L}}^2\geq{0}
    \end{split}
\end{equation}
where (\ref{eq:ProofProp1.2}) is equivalent to (\ref{eq:Lips}). Thus, (\ref{eq:Lips}) is the sufficient and necessary conditions for $z(\bm{x})$ is convex. From \cite[Section 3.1.3]{ConvexOpt}, it follows that $z(\bm{x})$ is convex if and only if, for $\forall{\bm{q}^g,\bm{y}}\in\mathbb{R}^N$:
\begin{equation}\label{eq:ProofProp1.1}
    \begin{split}
        z(\bm{q}^g)&\geq{z}(\bm{y})+\nabla{z}(\bm{y})^T(\bm{q}^g-\bm{y})\\
        &=\frac{1}{2}\bm{y}^T\bm{L}\bm{y}-f(\bm{y})+<\bm{L}\bm{y}-\nabla{f}(\bm{y}),\bm{q}^g-\bm{y}>\\
        &=-f(\bm{y})-<\nabla{f}(\bm{y}),\bm{q}^g-\bm{y}>\\
        &-\frac{1}{2}||\bm{q}^g-\bm{y}||_{\bm{L}}^2+\frac{1}{2}(\bm{q}^g)^T\bm{L}\bm{q}^g
    \end{split}
\end{equation}
where (\ref{eq:ProofProp1.1}) is equivalent to (\ref{eq:Upperbound}). Thus, (\ref{eq:Upperbound}) is also the sufficient and necessary conditions for $z(\bm{x})$ is convex. From the above analysis, it concludes this proof. Q.E.D.

\section*{{Appendix B}}
\noindent
\textit{Proof of Proposition 3:}

It follows from Proposition 2 that (\ref{eq:superlinear}) holds due to [P3.A]. As $g(\bm{q}^g)$ is an indicator function, we know $g(\bm{q}^g(k))=g(\bm{q}^{g\ast})=0$. Then (11) boils down to:
\begin{equation}\label{eq:fqg}
    f(\bm{q}^g(k))-f(\bm{q}^{g\ast})\leq\frac{2||\bm{q}^g(0)-\bm{q}^{g\ast}||_{\bm{L}}^2}{(k+1)^2}, \forall{k}\geq{1}
\end{equation}
We introduce $z(\bm{x})=f(\bm{x})-\frac{1}{2}\bm{x}^T\bm{H}\bm{x}$. 
From [P3.B], we have:
\begin{equation}\label{eq:z}
\begin{split}
    <\nabla{z}(\bm{q}^g)-\nabla{z}(\bm{y}),\bm{q}^g-\bm{y}>&=\\<\nabla{f}(\bm{q}^g)-\nabla{f}(\bm{y}),\bm{q}^g-\bm{y}>&-||\bm{q}^g-\bm{y}||_{\bm{H}}^2\\
    \geq||\bm{q}^g-\bm{y}||_{\bm{H}}^2&-||\bm{q}^g-\bm{y}||_{\bm{H}}^2=0
\end{split}
\end{equation}
By exploiting \cite[Theorem 2.1.3]{YN}, it follows from (\ref{eq:z}) that $z(\bm{x})$ is convex. For the convex function $z(\bm{x})$, for $\forall{\bm{q}^g,\bm{y}}\in\mathbb{R}^N$, it follows from \cite[Section 3.1.3]{ConvexOpt} that:
\begin{equation}
    \begin{split}
        z(\bm{q}^g)\geq{z}(\bm{y})+\nabla{z}(\bm{y})^T(\bm{q}^g-\bm{y})
    \end{split}
\end{equation}
It follows that:
\begin{equation}
    f(\bm{q}^g)\geq{f}(\bm{y})+<\nabla{f}(\bm{y}),\bm{q}^g-\bm{y}>+\frac{1}{2}||\bm{q}^g-\bm{y}||_{\bm{H}}^2
\end{equation}
Then, we have:
\begin{equation}\label{eq:a}
\begin{split}
    f(\bm{q}^g(k))&\geq{f}(\bm{q}^{g\ast})+<\nabla{f}(\bm{q}^{g\ast}),\bm{q}^g(k)-\bm{q}^{g\ast}>\\&+\frac{1}{2}||\bm{q}^g(k)-\bm{q}^{g\ast}||_{\bm{H}}^2
\end{split}
\end{equation}
From \cite[Section 4.2.3]{ConvexOpt}, it follows that:
\begin{equation}\label{eq:b}
    <\nabla{f}(\bm{q}^{g\ast}),\bm{q}^g(k)-\bm{q}^{g\ast}>\geq{0}
\end{equation}
Combining (\ref{eq:a}) and (\ref{eq:b}), we have:
\begin{equation}\label{eq:c}
    f(\bm{q}^g(k))-{f}(\bm{q}^{g\ast})\geq\frac{1}{2}||\bm{q}^g(k)-\bm{q}^{g\ast}||_{\bm{H}}^2
\end{equation}
From (\ref{eq:fqg}) and (\ref{eq:c}), we have:
\begin{equation}
    \frac{1}{2}||\bm{q}^g(k)-\bm{q}^{g\ast}||_{\bm{H}}^2\leq\frac{2||\bm{q}^g(0)-\bm{q}^{g\ast}||_{\bm{L}}^2}{(k+1)^2}, \forall{k}\geq{1}
\end{equation}
Then, we obtain:
\begin{equation}
    ||\bm{q}^g(k)-\bm{q}^{g\ast}||_{2}^2\leq\frac{4||\bm{q}^g(0)-\bm{q}^{g\ast}||_{\bm{L}}^2}{(k+1)^2\sigma_{\rm min}(\bm{H})}, \forall{k}\geq{1}
\end{equation}
Q.E.D.

\section*{{Appendix C}}
\noindent
\textit{Proof of Proposition 4:}
\begin{equation}\small
\begin{split}\label{eq:Proposition3:1}
    &m(\bm{q}^g(k))-m(\bm{\hat{q}}^{g\ast})\\
    &=m(\bm{q}^g(k))-f(\bm{{q}}^{g\ast})+f(\bm{{q}}^{g\ast})-m(\bm{\hat{q}}^{g\ast})\\
    &\leq{m}(\bm{q}^g(k))-f(\bm{{q}}^{g\ast})+\tau\\
    &={m}(\bm{q}^g(k))-f(\bm{q}^g(k))+f(\bm{q}^g(k))-f(\bm{{q}}^{g\ast})+\tau\\
    &\leq {m}(\bm{q}^g(k))-f(\bm{q}^g(k))+\frac{2||\bm{q}^g(0)-\bm{q}^{g\ast}||_{\bm{L}}^2}{(k+1)^2}+\tau
\end{split}
\end{equation}
where the first inequality follows by [P4.B], and the second inequality follows by [P4.C] and Proposition 2. With respect to ${m}(\bm{q}^g(k))-f(\bm{q}^g(k))$, we have:
\begin{equation}\small
\begin{split}\label{eq:Proposition3:2}
    &{m}(\bm{q}^g(k))-f(\bm{q}^g(k))\\
    &=\frac{1}{2}\big[||h(\bm{q}^g(k),\bm{d})-\bm{V}_r||_{\bm{\Phi}}^2-||h_l(\bm{q}^g(k),\bm{d})-\bm{V}_r||_{\bm{\Phi}}^2\big]\\
    &=\frac{1}{2}\big[||h(\bm{q}^g(k),\bm{d})-h_l(\bm{q}^g(k),\bm{d})+h_l(\bm{q}^g(k),\bm{d})-\bm{V}_r||_{\bm{\Phi}}^2\\
    &-||h_l(\bm{q}^g(k),\bm{d})-\bm{V}_r||_{\bm{\Phi}}^2\big]
    \\&\leq\frac{1}{2}\big[||h(\bm{q}^g(k),\bm{d})-h_l(\bm{q}^g(k),\bm{d})||_{\bm{\Phi}}^2+||h_l(\bm{q}^g(k),\bm{d})-\bm{V}_r||_{\bm{\Phi}}^2\\
    &-||h_l(\bm{q}^g(k),\bm{d})-\bm{V}_r||_{\bm{\Phi}}^2\big]\\
    &=\frac{1}{2}||h(\bm{q}^g(k),\bm{d})-h_l(\bm{q}^g(k),\bm{d})||_{\bm{\Phi}}^2
\end{split}
\end{equation}
Since $\Phi$ is a symmetric positive definite matrix, it follows by Cholesky decomposition that $\Phi$ can be expressed by the form $\Phi=\bm{E}^T\bm{E}$. Then, we have:
\begin{equation}\label{eq:Proposition3:3}
\begin{split}
    &||h(\bm{q}^g(k),\bm{d})-h_l(\bm{q}^g(k),\bm{d})||_{\bm{\Phi}}\\
    &=||~\bm{E}[h(\bm{q}^g(k),\bm{d})-h_l(\bm{q}^g(k),\bm{d})]~||_2\\
    &\leq||\bm{E}||_2||h(\bm{q}^g(k),\bm{d})-h_l(\bm{q}^g(k),\bm{d})||_2\\
    &=||\bm{E}||_2\delta
\end{split}
\end{equation}
Combining (\ref{eq:Proposition3:1})-(\ref{eq:Proposition3:3}), it follows that (\ref{eq:Stability}) holds. Q.E.D.

\section*{{Appendix D}}
\noindent
\textit{Proof of Proposition 5:}

From (\ref{eq:ApproximationModel})-(\ref{eq:pL}), $\bm{q}^{g}(k)=p_{\bm{L}}(\bm{y}(k))$ can be represented by:
\begin{equation}\label{eq:qg}\small
    \bm{q}^g(k)=\argmin_{\bm{\underline{q}}^g\leq\bm{q}^g\leq\bm{\overline{q}}^g}<\nabla{f}(\bm{y}(k)),\bm{q}^g-\bm{y}(k)>+\frac{1}{2}||\bm{q}^g-\bm{y}(k)||_{\bm{L}}^2
\end{equation}
For the diagonal positive definite matrix $\bm{L}$, (\ref{eq:qg}) is equal to:
\begin{equation}\label{eq:Proposition3}\small
    \bm{q}^g(k)
    =\argmin_{\bm{\underline{q}}^g\leq\bm{q}^g\leq\bm{\overline{q}}^g}\sum_{i=1}^{N}\Big\{\frac{\partial{f}(\bm{y}(k))}{\partial{y}_i(k)}[q_i^g-y_i(k)]+\frac{L_i}{2}[q_i^g-y_i(k)]^2\Big\}
\end{equation}
It is clear that both the objective and constraint in (\ref{eq:Proposition3}) are decomposable, thus, for any $i\in\mathcal{N}$.$q_i^g(k)$ can be solved by:
\begin{equation}\label{eq:qi}
    {q}_i^g(k)
    =\argmin_{{\underline{q}}_i^g\leq{q}_i^g\leq{\overline{q}}_i^g}\frac{\partial{f}(\bm{y}(k))}{\partial{y}_i(k)}[q_i^g-y_i(k)]+\frac{L_i}{2}[q_i^g-y_i(k)]^2
\end{equation}
Note that ${q}_i^g(k)$ is a scalar, (\ref{eq:qi}) is equivalently solved by (\ref{eq:LocalImple}). 
In this case, $\bm{q}^g(k)$  can be represented by (\ref{eq:RewrittenS1}). Q.E.D.


\end{document}